\documentclass[DIV=calc,paper=a4,fontsize=11pt,twocolumn]{scrartcl} 

\usepackage[english]{babel}
\usepackage[protrusion=true,expansion=true]{microtype}
\usepackage{amsmath,amsfonts,amsthm}
\usepackage[final]{graphicx}
\usepackage{xcolor}
\usepackage[normal,small,hypcap,up,labelfont=bf,textfont=it]{caption}
\usepackage{epstopdf}
\usepackage{subfig}
\usepackage{booktabs}
\usepackage{fix-cm}
\usepackage{amssymb,amsfonts}
\usepackage{bbm}
\usepackage{pstricks}
\usepackage{cite}
\usepackage[utf8]{inputenc}
\usepackage[perpage,symbol*]{footmisc}
\usepackage[varg]{txfonts}
\usepackage{balance}
\usepackage{fancyhdr}

\usepackage[pdfencoding=auto]{hyperref}
\usepackage{verbatim}
\usepackage{fontenc}
\usepackage{cuted}

\theoremstyle{definition}

\theoremstyle{plain}

\DeclareCaptionFont{mycolor}{\color[HTML]{0000FF}}
\usepackage{sectsty}                                                    
\allsectionsfont{
\color[HTML]{31ADF3}\usefont{OT1}{phv}{b}{n}
}

\sectionfont{
\color[HTML]{31ADF3}\usefont{OT1}{phv}{b}{n}
}

\usepackage{fancyhdr}                                               
\usepackage{lettrine}
\newcommand{\initial}[1]{%
\lettrine[lines=3,lhang=0.3,nindent=0em]{
\color[HTML]{31ADF3}
{\textsf{#1}}}{}}

\usepackage{titling}                                                            

\newcommand{\HorRule}{\color[HTML]{31ADF3}
\rule{\linewidth}{1pt}%
}

\pretitle{\vspace{-30pt} \begin{flushleft} \HorRule
\fontsize{34}{34} \usefont{OT1}{phv}{b}{n} \color[HTML]{31ADF3} \selectfont
}

\newcommand{\braket}[2]{\langle  #1|#2\rangle}       
\newcommand*{\ket}[1]{|#1\rangle}
\newcommand*{\bra}[1]{\langle #1|}

\title{Weak Measurement and Two-State-Vector Formalism: Deficit of Momentum Transfer in Scattering Processes}     
\posttitle{\par\end{flushleft}\vskip 0.5em}

\preauthor{\begin{flushleft}\large \lineskip 0.5em \usefont{OT1}{phv}{b}{sl} \color[HTML]{31ADF3}}
\author{Chariton Aris Chatzidimitriou-Dreismann\\[8pt]}
\postauthor{\footnotesize \usefont{OT1}{phv}{m}{sl}
\color[HTML]{000000} Institute of Chemistry (C2),
Faculty II, Technical University of Berlin, D-10623 Berlin,
Germany \\E-mail: \href{mailto:dreismann@chem.tu-berlin.de}{dreismann@chem.tu-berlin.de}\\[10pt]        
\scriptsize\usefont{OT1}{phv}{m}{n} \color[HTML]{31ADF3}{\textbf{Editors: \emph{Eliahu Cohen}
\& \emph{Tomer Shushi}} }\\[5pt]
\color[HTML]{000000}{Article history: Submitted on July 31,
2015; Accepted on October 20, 2016; Published on October 28, 2016.}
\par\end{flushleft}\HorRule}

\date{}                                                                             

\begin{document}
\maketitle
\thispagestyle{fancy}         
\initial{T}\textbf{he notions of weak measurement, weak value, and two-state-vector
formalism provide a new quantum-theoretical frame for extracting additional information from a
system in the limit of small disturbances to its state.
Here, we provide an application to the case of two-body scattering with one
body weakly interacting with an environment. The direct
connection to real scattering experiments is pointed out by making
contact with the field of impulsive incoherent neutron scattering
from molecules and condensed systems. In particular, we predict a new quantum effect in
neutron-atom collisions, namely an observable momentum transfer
deficit; or equivalently, a reduction of effective mass below that of the free scattering atom. Two
corroborative experimental findings are shortly presented.
Implications for current and further experiments are mentioned.
An interpretation of this effect and the associated experimental results within conventional theory is currently unavailable.\\
Quanta 2016; 5: 61--84.}

\begin{figure}[b!]
\rule{245 pt}{0.5 pt}\\[3pt]
\raisebox{-0.2\height}{\includegraphics[width=5mm]{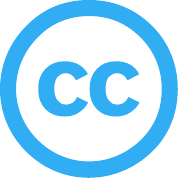}}\raisebox{-0.2\height}{\includegraphics[width=5mm]{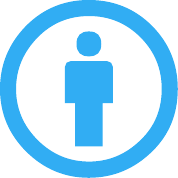}}
\footnotesize{This is an open access article distributed under the
terms of the Creative Commons Attribution License
\href{http://creativecommons.org/licenses/by/3.0/}{CC-BY-3.0},
which permits unrestricted use, distribution, and reproduction in
any medium, provided the original author and source are credited.}
\end{figure}

\section{\label{sec:1}Introduction}

Time symmetry, and the associated microscopic reversibility, is widely acknowledged to be a
fundamental property of the basic physical laws of classical and  quantum mechanics
\cite{Feynman1965}. In contrast, the Second Law of thermodynamics, although also being considered
fundamental, is inconsistent with this symmetry and provides an
\emph{arrow of time} associated with
entropy production appearing in isolated many-body systems. However, thermodynamics  and
statistical mechanics (classical or quantum), and the associated thermodynamic limit
(when the particle number $N \rightarrow \infty$) are not under consideration in this paper. Instead,
here we are particularly interested in (few-body) quantum dynamics
in connection with  quantum measurements
\cite{Wheeler1983,Wiseman2009,Nielsen2000}, in the realm of \emph{non-relativistic} quantum mechanics. In
particular, the emphasis will be
on an experimental application in the frame of
\emph{incoherent scattering} experiments. (An explanation from first principles of
\emph{coherent} versus \emph{incoherent}, being also helpful for the
neutron-proton scattering experiments considered below,
was presented by Feynman in \cite[Section 3-3]{Feynman1965}.)

In this context, the irreversible character of the standard (i.e. projective, strong)
measurements should be mentioned. In contrast  with the unitary (and thus
time-symmetric) time evolution generated by the Schr\"odinger equation, a strong (projective)
measurement \cite{Wheeler1983,Wiseman2009,Nielsen2000}
causes a reduction of the state of the measured system,
constituting an irreversible (i.e. time-oriented) process.

Environmentally induced decoherence (see e.g. the textbooks \cite{Nielsen2000,Schlosshauer})
breaks time symmetry of the equations of motion and thus provides an explanation for the
appearance of time-oriented quantum processes in open quantum systems. Nowadays decoherence
plays a crucial role in many fields of physics, chemistry and molecular biology, in quantum
information theory   and also in the currently emerging
\emph{quantum information technologies}, e.g. quantum computing,
communication and coding \cite{Nielsen2000}.

Novel insights into the  measurement problem of quantum mechanics, and
hence also into the foundations  of quantum theory, are provided by the
seminal work by Aharonov and collaborators,  commonly known as theory of \emph{weak measurements}
and/or \emph{weak values}; see e.g. the textbook \cite{Aharonov2005} and
the review article  \cite{Kofman2012}. Elegant introductory presentations are given in
\cite{Tamir2013,Svensson2013,Dressel2014}.
For a very recent discussion of the basic physical aspects,
see \cite{Aharonov2014}.

The outline of this article is as follows:
In Section~\ref{sec:2}, we begin by briefly reviewing  some  basic elements of the aforementioned theoretical frame,
since it plays a central role in the physical context of the present paper.
Section~\ref{sec:3} presents a few theoretical papers that motivated the application of the weak measurement theory
to the scattering topic being under consideration in this paper.
In Section~\ref{sec:4} are presented basic elements of the theory  of impulsive
scattering and the specific experimental method employed to a class of neutron scattering
experiments (as those presented in Section~\ref{sec:6}).
Section~\ref{sec:5}, which is the main theoretical part of the paper, investigates elementary scattering  in
the light of weak measurement and \emph{two-state-vector formalism}
\cite{Aharonov2005,Tamir2013,Svensson2013,Aharonov2014}.
The main theoretical result is the effect of
\emph{momentum transfer deficit} in impulsive collisions, and associated aspects of it (e.g. the
anomalous \emph{reduction of effective mass} of the scattering particle, or an \emph{increased
energy transfer}).
Two concrete neutron scattering
experiments  demonstrating the experimental applicability of the
weak measurement  theory, as well as its novelty, are presented in Section~\ref{sec:6}.
Finally, Section~\ref{sec:7} provides additional remarks to the main results and a discussion.

Here, it may be helpful to emphasize two particularly relevant points.
First, what we directly read out in a measurement of an experiment are not (values of)
the dynamical variables of the observed system, but the outcomes of
the measuring devices of the employed instrument (see subsection~\ref{sec:7.1}). To these outcomes belongs the measured data (in Section~\ref{sec:6})
that exhibits the peculiar phenomenon of momentum transfer deficit.
It will be demonstrated that the latter contradicts every prediction
of conventional theory even qualitatively, but  it can be
interpreted straightforwardly within  the theory of weak values and two-state-vector formalism.

Second, with respect to
an important point raised by Vaidman \cite{VaidmanComment2014}, we would like to
stress the following special feature of our investigation.
Until now, weak values are measured with a different degree of freedom of
the \emph{same}  quantum system. Here, however, a weak value is observed with an
\emph{external}   quantum system.
Hence, in the present investigation, the
weak value appears due to interference of a quantum entangled wave,
thus having no analog in classical wave interference \cite{VaidmanComment2014}.

\section{\label{sec:2}On weak measurement, post-selection and two-state-vector formalism}

Weak measurement  is unique in measuring non-commuting operators and revealing new
counter-intuitive effects predicted by the new theory of weak values and  two-state-vector formalism
\cite{Aharonov2014,Aharonov1990,Aharonov-PhysToday}.
The main aim of this article is not to further develop and/or extend this theory, but to point out
certain  new (and experimentally observable) features of elementary scattering processes predicted
within the theoretical frame of  weak values and two-state-vector formalism. Especially,
\emph{incoherent} scattering of single (massive) particles, like
neutrons or electrons, from nuclei and/or atoms is investigated.

As the starting point of the aforementioned theory one usually considers the paper
\cite{AAV1988} by Aharonov, Albert and Vaidman, and the earlier paper \cite{ABL1964} by Aharonov,
Bergmann and Lebowitz; see also \cite{Duck1989} for a clarifying discussion. Here let us shortly
mention a few results needed in the following.

According to the standard theory (of ideal, projective von Neumann
measurements \cite{vonNeumann}),
the final state of the
system after a measurement becomes an eigenstate of the measured observable. This usually
disturbs the state of the system. On the other hand, by coupling a measuring device to a
system sufficiently weakly,  it
may be possible to read out certain information while limiting the
disturbance induced by the measurement to the system. As Aharonov and collaborators originally
proposed  \cite{AAV1988,ABL1964},
one may achieve new  physical insights  when one furthermore
post-selects on a particular outcome
of the experiment. In this case the eigenvalues of the measured observable
are no longer the relevant quantities; rather the measuring device consistently indicates the {\em
weak value} given by \cite{AAV1988,ABL1964}
\begin{equation}
\label{AAV-formula}
A_w \equiv (\hat{A})_w = \frac{\bra{\psi_f} \hat{A}
\ket{\psi_i}}{\braket{\psi_f}{\psi_i}},
\end{equation}
where $\hat{A}$ is the operator of the quantity being measured,
$\ket{\psi_i}$ is the initial (pre-selected) state of the system, and $\ket{\psi_f}$ is the post-selected
state  after the weak measurement. Note that the number $A_w$ may be
complex.

The significance of this formula is as follows. Let us couple a measuring device whose pointer has
position coordinate $\hat{q}$ to the system's dynamical variable  $\hat{A}$, and subsequently
measure its conjugated momentum $\hat{p}$. The coupling interaction is taken to be the standard
von Neumann measurement interaction \cite{vonNeumann}
\begin{equation}
\hat{H}_\textrm{vN}   = - g(t) \, \hat{q} \otimes \hat{A}.
\label{interaction}
\end{equation}
The coupling factor $g(t)$ is assumed to be appropriately small and to have a
compact support at the time of the (impulsive) measurement; e.g. it may be
proportional to a delta function. Since it is time-dependent, the
complete Hamiltonian of the system and the measuring device
represents an \emph{open quantum system}.
Consequently, in the situations
contemplated here the time dependence of the coupling constant
implies that energy conservation  need not apply.

The mean value $\langle \hat{p} \rangle_f$ of the pointer momentum after the measurement is given
by \cite{AAV1988}
\begin{equation}
\langle \hat{p} \rangle_f - \langle \hat{p} \rangle_i =  +  g\, \textrm{Re} [A_w],
\label{ReWV}
\end{equation}
where $\langle \hat{p} \rangle_i$ is the corresponding value before measurement, $\textrm{Re}$ denotes the
real part and $g =\int g(t)dt$.

Formula (\ref{AAV-formula}) implies that, if the initial state $\ket{\psi_i}$ is an eigenstate
of a measurement operator
$\hat{A}$,
then the weak value post-conditioned on that eigenstate is the
same as the classical (strong) measurement result. When there is a definite outcome, therefore,
strong and weak measurements agree. However, a weak measurement  can yield values outside the  range of
measurement results predicted by conventional theory \cite{AAV1988}.

A weak value can also be complex, with an imaginary part corresponding  to the pointer position. In fact,
the mean of the pointer position after measurement is given by
\begin{equation}
\label{ImWV}
\langle \hat{q} \rangle_f = -   2gv_q\ \textrm{Im} [A_w],
\end{equation}
where $\textrm{Im}$ denotes the imaginary part and $v_q$ is the variance in the initial pointer spatial
position \cite{AAV1988}, assuming that $\langle \hat{q} \rangle_i =0$ before measurement.

In the simplest case where there is just one observable $\hat{A}$, we assume the evolution from
$\ket{\psi_i}$ to the point where $\hat{A}$ is measured is given by $\hat{U}$, and from this point
to the post-selection the evolution is given by $\hat{V}$. Then we can rewrite formula (\ref{AAV-formula})
as:
\begin{equation}
\label{fullAAV}
A_w \equiv (\hat{A})_w
=\frac{\bra{\psi_f}\hat{V}\hat{A}\hat{U}\ket{\psi_i}}{\bra{\psi_f}\hat{V}\hat{U}\ket{\psi_i}}
\end{equation}
and the expressions (\ref{ReWV}) and (\ref{ImWV}) characterizing  the pointer shift remain valid.

The fact that one only
\emph{sufficiently weakly} disturbs the system in making weak measurements
implies that one can in principle measure different (also non-commuting) dynamical
variables in succession. This theoretical observation has led to a great number of experimental
applications and discovery of several new effects; see e.g.
\cite{Kwiat2008,Jordan2009,Lundeen2011,Steinberg2011,Vaidman2013}.

The generalization of the concept of weak value to a system described by a density operator $\hat{\rho}_i$
(for its initial state) was first considered by Wiseman \cite{Wiseman2002}:
\begin{equation}
(\hat{A})_w =
\frac{\bra{\psi_f} \hat{A} \hat{\rho}_i\ket{\psi_f}}{\bra{\psi_f}\hat{\rho}_i\ket{\psi_f}};
\end{equation}
cf. also \cite{DiLorenzo2008}.

Furthermore, it may be noted that there is (steadily growing) evidence that the weak value formalism can
be extended beyond the weak coupling regime of the original framework \cite{AAV1988}; see
e.g.~\cite{Kofman2012,Zhu2011,Pang2012}.  Moreover, recently it has been shown that
measurements on macroscopic systems are  weak measurements \cite{Gisin2014}, which also provided
new physical inside to quantum non-locality  \cite{Gisin2014,Gisin2015}.

Technical and experimental merits, and associated advantages (or disadvantages) to precision
metrology, of weak values-based techniques have been discussed in various works, e.g. in
\cite{JordanPRX,Walmsley2015}.

Although a discussion and/or analysis of the physical interpretation and/or meaning of the concepts
of weak values and two-state-vector formalism are beyond the scope of this paper, a few related
short remarks may be in order.

Nowadays, the importance of weak values and associated weak measurement as a \emph{practical tool} for describing new
experiments and extract new information from  experiments seems to be broadly acknowledged;
cf.~the discussions in \cite{Dressel2014}. However, various criticisms have been provided
which deny or question the novelty of these quantum mechanical concepts. For example, in
\cite{Ferrie-Combes} it is claimed that weak value is not an inherently quantum concept but rather a
purely statistical feature of pre- and post-selection with disturbance; and furthermore, that
classical correlation alone supplies the surprising \emph{anomalies} revealed with weak values.
For some concise comments stressing quite the opposite, see \cite{VaidmanComment2014,Romito2015}.

For the interpretation of the experimental findings presented below, it is relevant
that weak values are novel quantum interference phenomena, having no classical analogue,
in which post-selection plays a crucial role;
cf.~\cite{VaidmanComment2014,Romito2015,ParksGray2012,Pati2014,Dressel2015,Qin2015}.

\section{\label{sec:3}Some previous results and motivation}

This paper  concerns the measurement of \emph{momentum} and \emph{energy} transfers
in real scattering experiments, the corresponding predictions of conventional theory, and  a new
prediction based on the formalism of weak measurement and two-state-vector formalism.
The latter point may be best illustrated and motivated by
referring  directly to some of the intriguing results presented in two recent papers
\cite{Vaidman2013,Aharonov2013}. Additionally, we mention here certain interpretational issues
concerning the theoretical status of weak values \cite{VaidmanComment2014}, because they contributed to
the motivation  of the present investigation  and  facilitate the interpretation of the obtained results.

First, let us refer to a surprising theoretical prediction derived by Aharonov et al. in
\cite{Aharonov2013}, which may be shortly described (with some simplifications) as follows.

A photon beam enters a device similar to a usual Mach--Zehnder interferometer, with the
exception that one reflecting mirror is sufficiently small (say, a nanoscopic object $M$)
for its momentum distribution to be detectable
by a suitable non-demolition measurement
\cite{Scully1997}. The two identical beam splitters of the Mach--Zehnder interferometer
have nonequal reflectivity $r$ and transmissivity $t$ (both real, with $r^2 + t^2 = 1$),
say $r > t$.  Now one is interested in the momentum kicks given to the mirror $M$
caused by the photon reflection on $M$ inside the interferometer, but only for
photons emerging toward one of the two detectors (D$_2$ in Fig.~2 of \cite{Aharonov2013}).
The latter condition is a specific post-selection.   The effect of the photons emerging toward
the second detector is discarded.

A  straightforward calculation \cite{Aharonov2013}  shows  the following astonishing feature.
Although the post-selected photons  collide
(as all photons do, of course) with the mirror $M$ only
from the \emph{inside} of the Mach--Zehnder interferometer, they do not push $M$ outwards, but rather they somehow succeed
to \emph{pull it in}.  It is obvious that this result cannot have any conventional theoretical
interpretation.  As Aharonov et al.~put it:
\begin{quote}
This is realized by a superposition of
giving the mirror zero momentum and positive momentum---the superposition results in the mirror gaining 
negative momentum. \cite{Aharonov2013}
\end{quote}

Another paper by Vaidman and collaborators \cite{Vaidman2013} presents
\emph{experimental} and theoretical results of optical measurements with a special interferometer
being a combination of two Mach--Zehnder interferometers; essentially, a
second Mach--Zehnder interferometer is put on the place of one of the two
reflecting mirrors of the first Mach--Zehnder interferometer. The whole construction consists of 5 mirrors and 4 beam
splitters (see \cite[Figs.~2(b) and 3]{Vaidman2013}). A continuous laser beam enters the
interferometer.

Furthermore, all 5 mirrors are  placed on piezoelectrically driven mirror mounts and are weakly
vibrating at \emph{different} frequencies (say $f_i$) and produce very slight rotational motions,
which correspondingly produce slight deflections of the photon beam. The vertical displacements of
the beam due to the vibrations of the mirrors are significantly smaller than the width of the beam,
and the change in the optical path length is much smaller than the wavelength. The photon beam
coming out  of the interferometer (toward a specific detector) is measured and Fourier analyzed.
It is natural to expect that  the specific frequencies appearing in the Fourier power spectrum
should be those of the mirrors  the photons bounce off.

The reported experimental results are quite unexpected. For example, in one specific setup (see \cite[Fig.~3]{Vaidman2013}),
\emph{three} frequencies  (i.e.~those of mirrors A, B and C) appear in the
measured power spectrum instead of the naturally expected one frequency (i.e.~that
of mirror C). Moreover, it appears that the past of the photons is not represented by continuous
trajectories.

However, these striking results have a simple explanation in the framework of two-state-vector formalism.
As Vaidman et al.~propose, the intuitive picture which allows
us to understand the experimental findings is provided by the
time-symmetric two-state-vector formalism. Here each photon observed by detector
($D$ in \cite[Fig.~3]{Vaidman2013})  is described
by the backward-evolving quantum state $\bra{\psi_f}$
post-selected at $D$, in addition to the standard,
forward-evolving wave function $\ket{\psi_i}$
pre-selected at the  photon source.
The formulas of two-state-vector formalism imply that a photon can have a local observable
effect only if both the forward- and backward-evolving quantum waves are non-vanishing at the
considered  location; see \cite[Fig.~3]{Vaidman2013}, in which both forward-in-time and
backward-in-time paths of traveling photons are shown.

In other terms, one may say that each photon was present in a specific position (i.e.
at a specific mirror) only when both forward- and
backward-evolving quantum wave functions do not vanish at that position---this happens only at
three of the five mirrors.
This provides an explanation from first principles of the observed power spectrum in the frame of two-state-vector formalism, thus
also illustrating the predictive power of the theory.

Responding to certain claims that question the novelty of weak values,
Vaidman states the following:
\begin{quote}
The weak value shifts exist if measured or not, so the weak value
is not defined by the statistics of measurement outcomes. The
statistical analysis (performed after the post-selection) can just
reveal the pre-existing weak values. $\left[\ldots\right]$\\
The concept of weak value arises due to wave interference and has
no analog in classical statistics. Moreover, if weak values are observed with
external systems (and not with a different degree of freedom of the
observed system as it has been done until now) then the weak value
appears due to interference of a quantum entangled wave and it has
no analog in classical wave interference too. Therefore, weak value
is a genuinely quantum concept \cite{VaidmanComment2014}.
\end{quote}

Concluding the above short remarks, one may say that the new
insights and predictions made possible within the theoretical frame
of weak measurement, weak values and two-state-vector formalism are not
limited to interpretational issues only. The revised intuitions can then
lead one to find  novel quantum effects that can be measured in real experiments.

\section{\label{sec:4}Elementary remarks on impulsive scattering}

Throughout this paper  \emph{non-relativistic} quantum mechanics is considered.
In this section, we give an outline of basic elements of  impulsive scattering experiments. In
particular, we consider \emph{incoherent inelastic neutron scattering}
from condensed matter (see
the textbooks  \cite{Squires, Lovesey})
and its specification to high momentum transfers, known  as
\emph{neutron Compton scattering} or \emph{deep-inelastic neutron scattering};
cf.~\cite{Watson,Tietje2011,Sears}. (The neutron-nucleus collision is elastic; ``inelastic" simply
refers to the decreased kinetic energy of the neutron's final state; see Eq.~(\ref{EnergyLoss}).)
The initial energy of the neutrons under consideration is roughly in
the range of meV (for \emph{cold} neutrons) until about 100 eV (for \emph{epithermal} neutrons). The de
Broglie wavelength of the neutrons is roughly in the range \mbox{0.1--10 \AA}.  In two-body
(i.e. neutron-nucleus) collisions, the scattering is isotropic in the center-of-mass system, which
is called $s$-wave scattering. This is because the range of the nuclear strong force, and also the
dimension of nuclei, are some orders of magnitude smaller than neutron's de Broglie wavelength.

The usual experimental method employed at pulsed neutron sources
is \emph{time-of-flight}; see below.

A similar outline applies also to \emph{electron} Compton-like
quasielastic scattering from atoms (or molecules) in the gas phase;
see e.g.~\cite{Bonham2009}. In a conventional electron scattering
spectrometer, instead of time-of-flight, the experimental method uses the
deflection of the scattered electron in an electric field, in order
to determine the electron's final energy. However, a new generation
of instruments provides pulsed electron beams and applies the time-of-flight
analysis; cf.~\cite{BonhamReview}.

\subsection{\label{sec:4.1}Scattering experimental setup -- What is measured}

\begin{figure}[ht]
\begin{center}
\includegraphics[width=85 mm]{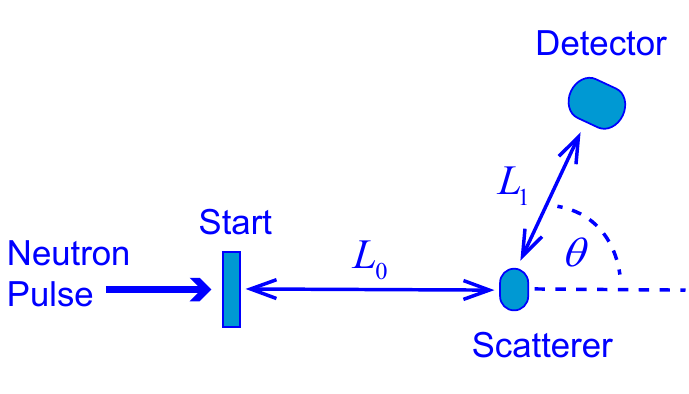}
\end{center}
\caption{\color[HTML]{0000FF}{\label{fig:1} Schematic of a time-of-flight scattering experiment.}}
\end{figure}

First, let us give an outline of a standard time-of-flight
scattering experiment; see Fig.~\ref{fig:1}.
A pulsed source of particles,  say neutrons, is used. A short pulse
reaches the monitor which triggers the measurement of time-of-flight. A neutron
scatters from the sample and reaches the detector, which stops the
time-of-flight measurement. For a measured time-of-flight value $t_\textrm{TOF}$ holds
\begin{equation}
t_\textrm{TOF}=\frac{L_0}{v_0}+\frac{L_1^\theta}{v_1} + t_0 .
\label{TOF}
\end{equation}
Here, $L_0$ is the distance between source and sample, $L_1^\theta$
is the sample--detector distance; the detector is positioned at the
scattering angle $\theta$; $v_0$ and $v_1$ describe the neutron's
velocities before and after scattering, respectively; $t_0$ is a
(usually small) time offset arising due to electronic delays. In a
so-called direct geometry spectrometer, the energy of the incident
neutrons is chosen as constant, $E_0=$constant.

In a general scattering experiment the scattering intensity is
measured as a function of the neutron energy transfer (or: energy loss)
\begin{eqnarray}
E  &=&  E_0- E_1 =
\frac{1}{2}m v_0^2 - \frac{1}{2}m v_1^2  \nonumber \\
&=&    \hbar \omega = \frac{(\hbar k_0)^2}{2m} - \frac{(\hbar k_1)^2}{2m}
\label{EnergyLoss}
\end{eqnarray}
($m$: neutron mass) due to the neutron-atom collision, and the corresponding neutron momentum
transfer on the struck atom
\begin{equation}
\hbar {\mathbf K}  =\hbar {\mathbf k}_0 - \hbar {\mathbf k}_1 ,
\label{K-vector}
\end{equation}
where
\begin{equation}
| {\mathbf K} |= K =\sqrt{k_0^2+k_1^2 - 2k_0k_1\cos \theta} .
\label{K}
\end{equation}
The subscripts 0 and 1 refer to quantities before and after the neutron-atom
collision, respectively. Since the scattering under consideration is \emph{incoherent} \cite{Feynman1965}, it
causes exchange of momentum between a neutron and (the nucleus of) a single  atom.

In the standard theory of elastic scattering from a free atom with mass
$M$ and zero initial momentum, \mbox{$\langle P \rangle =0$}, conservation
of kinetic energy and momentum in an elastic neutron-atom collision
yield the simple kinematic
relation  \cite{Tietje2011}
\begin{equation}
\frac{k_1}{k_0}= \frac{\cos\theta + \sqrt{(M/m)^2
- \sin^2\theta}} {M/m + 1} .
\label{k1k0}
\end{equation}
More precisely, this relation holds for the center-of-gravity  of the measured intensity peak.

As several experimental details play a crucial role in the
theoretical framework under consideration (since they concern pre-selection
and post-selection), the following facts should be pointed out.

$(a)$  From the measured time-of-flight value (\ref{TOF}), but without using the actual value of scattering
angle $\theta$, follows the experimental value of $k_1 =  | {\mathbf k}_1|$,  and consequently of
energy transfer $E=\hbar \omega$; see Eq.~\ref{EnergyLoss}.

$(b)$ Momentum transfer $\hbar {\mathbf K}$, Eq.~(\ref{K}), is determined from both
the scattering angle $\theta$ and the time-of-flight value.

Summarizing, from each value of $t_\textrm{TOF}$ measured with the
detector at $\theta$, the associated transfers of momentum ($\hbar K
$) and energy ($E=\hbar \omega$)   of the neutron to the struck
particle are uniquely determined.  Hence a specific detector
measures
one specific trajectory in the whole $K$--$E$~plane
only. Clearly, this  is related  to the \emph{post-selection} of the theory of weak values
and two-state-vector formalism. To the pre-selection belong $E_0$ and the initial
velocity of the sample which in most cases is at rest.

The scattering process produces an outgoing three-dimensional wave
for the entangled atom-neutron system, which is isotropic in the
center-of-mass system \cite{Squires,Lovesey}. The process  is called
$s$-wave scattering; see also below.

In the typical neutron
scattering experiment, the recoiling atoms of the sample are not measured. However, in other
scattering technics (as e.g. employed in high-energy investigations), the detectors can measure
most of the particles participating in the collision process.

A specific
neutron detector measures the time-of-flight of incoming neutrons at its position, and so effectuates a
reduction of the scattering wave by post-selecting specific components of this wave.
Thus, according to Eqs.~(\ref{EnergyLoss}) and (\ref{K}),
the detector's  instrumental parameters shown in Eq.~(\ref{TOF}) (denoted by $\{\textrm{IP}\}$) and a
confined range of time-of-flight values  determine the corresponding scattering intensity
$I(\textbf{K}, E)$, i.e.
$$
I(\textrm{TOF}, \{\textrm{IP}\}) \ \  \Rightarrow \ \  I(\textbf{K}, E)
$$
(and, furthermore, the dynamical structure factor $S(\textbf{K},\omega)$;
see below) of the scattering system. This constitutes a partial result of the whole measurement,
since the instrument may have many detectors at various angles $\theta$, and also measure a broad
range of time-of-flight values.

\subsubsection{\label{sec:4.1.1}On momentum and energy conservation -- Two-body collision}

When scattering takes place and the detector registers a scattered neutron, the impinging neutron
causes the aforementioned momentum transfer $ +\hbar \textbf{K} $ to the atom.
Due to  momentum conservation, it follows that
the neutron receives the opposite momentum $- \hbar \textbf{K} $.

The elastic collision of a neutron and a (free)  atom with mass $M$
and initial momentum $\textbf{P}$ results in the neutron's lost
energy $E\equiv\hbar \omega$ being transferred to the struck atom:
\begin{eqnarray}
E= E_0- E_1 = \hbar \omega & = &
\frac {(\hbar \mathbf K + \mathbf P )^2}{2M} -\frac{P^2}{2M}  \nonumber \\
& = & \frac {(\hbar K)^2}{2M} +\frac{\hbar{\mathbf K\cdot \mathbf P}}{M} .
\label{E-conserv}
\end{eqnarray}
This equation represents energy conservation.
As above, $\hbar {\mathbf K  }$ is the momentum transfer  from the
neutron to the struck atom.

\begin{figure}[ht]
\begin{center}
\includegraphics[width=85 mm]{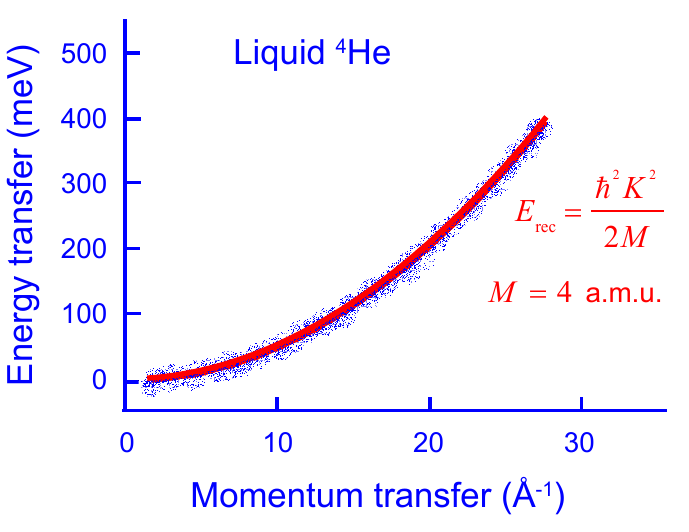}
\end{center}
\caption{\color[HTML]{0000FF}{\label{fig:2} Schematic representation of measured dynamic structure factor $S(K,E)$ of liquid helium \cite{Glyde}.
The red line is the calculated recoil parabola, Eq.~(\ref{recoil}), for the mass of $^4$He, shown as a guide to
the eye. The white-blue ribbon  around the recoil parabola represent data points measured with
the time-of-flight spectrometer ARCS \cite{ARCS}  (Adapted from \cite[Fig.~1]{Glyde}.)}}
\end{figure}

The first term in the right-hand side   defines the so-called recoil energy,
\begin{equation}
E_\textrm{rec}=\hbar\omega_\textrm{rec} = \frac{(\hbar K)^2}{2M} ,
\label{recoil}
\end{equation}
which  represents the kinetic energy of a recoiling  atom being
initially at rest.
In the latter case, $\langle P \rangle=0$ and thus one may write
\begin{equation}
\langle E \rangle = \frac{ \hbar^2 \langle K \rangle^2 }{2M} \equiv  \bar{E}_\textrm{rec} ,
\label{E-conserv2}
\end{equation}
which holds at the peak center.
Thus incoherent scattering from (a gaseous sample of) such atoms leads to a experimental recoil peak centered
at energy transfer  $\bar{E}_{\textrm{rec}}$, and exhibiting a
width being caused by the term $\hbar {\mathbf K \cdot \mathbf P}/M$ which represents
\emph{Doppler} broadening. Both entities are nicely illustrated in Fig.~\ref{fig:2}, which shows data of
incoherent inelastic (Compton) scattering from $^4$He atoms in  the liquid phase; see \cite{Glyde}
for details.

Now consider the formal structure of the Doppler-term
\begin{equation}
E_\textrm{Doppler} = \frac{\hbar{\mathbf K\cdot \mathbf P} }{ M} \equiv \frac{\hbar K P_\|}{ M} ,
\label{Doppler}
\end{equation}
where $ P_\| $ is the component of the atomic momentum parallel to $\textbf{K}$. For
\emph{isotropic} systems (gases, liquids, amorphous solids) the specific direction determined by
$\textbf{K}$ becomes immaterial, and thus $P_\| $ means the projection along any direction.

\subsubsection{\label{sec:4.1.2}On basic theory -- Scattering interaction and  dynamic structure factor}

The basic  quantity to be experimentally determined in a neutron scattering experiment is the
partial differential cross-section \cite{Squires}
\begin{equation}
\frac{d^2\sigma_M}{d\Omega d\omega}  =
\frac{k_1}{k_0} b_M^2 S(\textbf{K},\omega) ,
\label{pdcs}
\end{equation}
which is proportional to the measured intensity $I(\textbf{K},\omega)$. $E=\hbar \omega$, and
$S(\textbf{K},\omega)$ is the dynamic structure factor of the scattering system \cite{Squires}.
Note that its  dependence on momentum \emph{transfer} $\hbar \textbf{K} $ (rather than on initial and
final momenta) is a specific feature of the first Born approximation;
$b_M$ is the so-called
scattering length, i.e. a nucleus-depending constant ($M$ indicating
the nucleus     as well as its mass) appearing in the empirically
introduced \emph{Fermi pseudo-potential}
\begin{equation}
V_M(r)= \frac{2\pi \hbar^2}{m} b_M \delta(r)
\label{Fermi}
\end{equation}
with $r$ being the relative distance between the two colliding particles.

As already pointed out in textbooks, the standard (or
conventional) theory of neutron scattering is based on  Fermi's
golden rule which is equivalent to the first Born approximation; both are
based on the formalism of first-order perturbation theory (see
e.g~\cite[p.~16]{Squires}). Moreover, it should be pointed out
that, in principle, this theory is inapplicable  to scattering by
the singular potential (\ref{Fermi}); the Fermi pseudo-potential is
a formal artifice defined to reproduce, in the first Born
approximation, what we know to be the correct behavior for simple
$s$-wave scattering (cf.~\cite[p.~11]{Lovesey}).

Consider a scattering system consisting of $N$ (identical, for simplicity) atoms.
The dynamic structure factor fulfills the relation
\begin{equation}
S({\mathbf K}, \omega) = \frac{1}{2\pi} \int_{-\infty}^{\infty} e^{-  \imath \omega \, t} F({\mathbf
K},t)\, dt ,
\label{S}
\end{equation}
where $F({\mathbf K},t)$ is the so-called \emph{intermediate correlation function} given by
\begin{equation}
F({\mathbf K},t)= \frac{1}{N}\sum_{j,k}^{N} \left\langle
e^{-  \imath {\mathbf K}\cdot \hat{{\mathbf r}}_j(0)}  e^{\imath {\mathbf K}\cdot \hat{{\mathbf r}}_k(t)}
\right\rangle
\label{F}
\end{equation}
with $\langle \ldots \rangle$ denoting a thermodynamic average \cite{Squires}. This function contains the
Heisenberg operators
\begin{equation}
\hat{ {\mathbf r}}(t)=\hat{U}^\dag(t) \hat{{\mathbf r}} \hat{U}(t) ,
\end{equation}
where $\hat{U}(t) $ is the unitary time evolution operator
\begin{equation}
\hat{U}(t)  = e^{-\frac{\imath}{\hbar} \hat{H}_c t}
\end{equation}
with $\hat{H}_c$ being the Hamiltonian of the complete $N$-body scattering system---not only of a single 
atom \cite{Squires,Lovesey}.

In the limit of sufficiently large momentum transfers holds the \emph{incoherent approximation}, in which
terms with $j\neq k$ are neglected
\begin{equation}
F({\mathbf K},t)= \left\langle
e^{-\imath {\mathbf K}\cdot \hat{{\mathbf r}}_j(0)} e^{\imath {\mathbf K}\cdot \hat{{\mathbf r}}_j(t)}
\right\rangle .
\label{F-incoh}
\end{equation}
Neutron-proton scattering is mainly incoherent also for \emph{small} $K$, due to the spin-flip
mechanism described by Feynman \cite{Feynman1965}, as mentioned in the Introduction.

The following point should be emphasized. Quantum  correlations---or even entanglement and/or
quantum discord---between
neutron and a scattering particle are absent in the whole conventional theory of neutron
scattering; see e.g.~the textbook \cite{Squires} and the review article \cite{Watson}. Moreover,
the notion of coherence length of the neutron plays no role in conventional neutron
\emph{scattering} theory. (Surely, this is not the case in neutron \emph{interferometry}.) This
is essentially the consequence of the first-order perturbation theoretical frame of conventional
theory.

\subsubsection{\label{sec:4.1.3}On impulsive scattering -- Impulse approximation}

The key approximation called \emph{impulse approximation} \cite{Watson}, applied in the so-called
neutron Compton scattering regime,  is the assumption that the collisional time $\tau$ of a
neutron with a nucleus is very short. Thus the atomic (nuclear) position $r_M(t)$ of an atom with
mass $M$ can be replaced by
\begin{equation}
{\mathbf r}_M(t) = {\mathbf r}_M(0) + t \frac{{\mathbf P}(0)}{M} ,
\label{t-short}
\end{equation}
where ${\mathbf P}(0)$ is the initial atomic momentum. This approximation assumes that the particle
travels freely over short enough times $0\leq t \leq \tau$,  and thus that its interaction with
other particles can be neglected.

In other words, it is assumed that the scattering particle recoils freely from the collision, with
interparticle interaction in the final state being negligible and the wave function of the
particle in its \emph{final} state being a \emph{plane wave} \cite{Watson,Sears}.

In a real experiment, the characteristic time $\tau$ for this collision is  extremely short, $\tau
\rightarrow 0$, since the neutron-nucleus potential $V_M$ is
essentially  a delta function, Eq.~(\ref{Fermi}).   An expression
for the
scattering time  in the impulse approximation is given by Sears,
\begin{equation}
\tau = \frac{M}{K \Delta P} ,
\end{equation}
where $\Delta P$ is  the width of the momentum distribution of the struck atom \cite{Watson,Sears}.
Depending on the specific experimental parameters (and mass of scattering atom),
$\tau$  lies
roughly in the femtosecond, or attosecond, time range \cite{Watson,Tietje2011,Sears}.

In the context of weak measurement and two-state-vector formalism,
it is important to observe that the basic Eqs.~(\ref{S},\ref{F}) depend
on dynamical variables of the scattering system only  \cite{Squires})
and no dynamical variable of the neutron---e.g. the parameter $b_M$ is not a dynamical variable but a c-number.

In the limit of sufficiently large momentum transfer, i.e. in
the \emph{impulse approximation},  conventional theory yields for the dynamic structure
factor the simple expression \cite{Watson,Tietje2011,Sears}
$$
S_M({\mathbf K}, \omega) =  \left\langle
\delta\left(\hbar\omega -\hbar \omega_\textrm{rec} -
\frac{\hbar{\mathbf K}\cdot{\mathbf P}}{M} \right)\right\rangle   \ \ \ \ \ \ \ \ \ \ \ \ \ \ \ \ \
$$
\begin{equation}
\ \ \ \ \ \ \ \ \ \ \ \ \  =   \int \! n_M({\mathbf P})\, \delta \left( \hbar \omega -
\hbar\omega_\textrm{rec} - \frac{\hbar {\mathbf K \cdot
\mathbf P}}{M} \right) dP ,
\label{S-IA}
\end{equation}
where $n({\mathbf P})$ is the classical distribution function of the atomic momenta in the initial
state and the delta function represents the aforementioned energy conservation (\ref{E-conserv}).
As the scatterer is initially at rest, it holds $\langle P \rangle =0$.

For the  case of one atom with momentum-space wave function $\Xi(P) $ (i.e. the Fourier transform
of the wave function $\Psi(r)$  in position space), this is given by
\cite{Watson,Tietje2011,Sears}
\begin{equation}
n_M(P)= | \Xi(P) |^2 .
\label{n-IA}
\end{equation}
This formula physically means that here  the measured signal (intensity, dynamical structure factor)
contains an \emph{incoherent} sum (superposition) over the classical probability distribution of
initial-state momenta---and not a coherent superposition of the scattering amplitudes before taking the absolute
square. In other terms, this expression contains only the
\emph{diagonal} part of the full density operator $| \Xi \rangle\langle \Xi |$, i.e.~all
non-diagonal terms are discarded. Also this weakness is a consequence of the
first-order perturbation theory underlying the impulse approximation  of conventional  theory.

A remarkably concise derivation of the main result, Eq.~(\ref{S-IA}), including a short
explanation of incoherent and  impulse approximations, is given by Sears in the
first three pages of \cite{Sears}.

\subsection{\label{sec:4.2}Conventional final-state effects, peak shift, and effective mass}

In real experiments, deviations from the impulse approximation may be observed. Here, we
will consider them within the frame of conventional theory.

The aforementioned energy conservation relation (\ref{E-conserv}) for a
two-body collision
holds exactly in the impulse approximation, but is
not completely fulfilled at  momentum and energy transfers of actual experiments, in which
so-called \emph{final-state effects} may become apparent. (In fact, this  term commonly includes
both \emph{initial and final} state effects; see \cite{Watson,Sears} and papers cited therein.)
These are caused by environmental forces on the struck particle, which affect both initial and
final states of it. Here, we will shortly discuss this effect in the frame of conventional theory.

If the scatterer is not completely free but \emph{partially bound} to other particles of its
environment, then the probe particle  scatters on
an  object with \emph{higher} effective mass, as the particles of
the environment have a certain mass; in other terms,  the scattering
particle becomes hindered by the environmental forces. The latter
may be due to some conventional mechanism of binding
(such as ionic or van der Waals forces; chemi- or physisorbtion).
However these forces can never cause  an increase of the particle's mobility, or equivalently, a
reduction of its  effective mass. The particle is \emph{dressed} by
certain environmental degrees of freedom, which is tantamount to a
small increase of its effective mass,
\begin{equation}
M_\textrm{eff} \geq M_\textrm{free} \equiv M .
\label{M-eff}
\end{equation}
The above remarks correspond to a well understood effect, often observed in scattering
from condensed systems; cf. \cite{Watson,Tietje2011,Sears}.

These illustrative considerations imply that relation (\ref{M-eff})
should be valid also in the  \emph{final} state of the struck
particle.  Moreover, it should hold under the conditions of neutron Compton scattering
as well as of incoherent inelastic neuron scattering.

This effect can be also shown  by referring to the aforementioned energy conservation relation,
see (\ref{E-conserv2}), here including a small term $E_\textrm{int} > 0$ describing the
atom-environment interaction:
\begin{equation}
\bar{E}  =
\frac {\hbar^2 \bar{K}^2}{2M}  + E_\textrm{int} ,
\label{E-conserv3}
\end{equation}
where $\bar{E}$ and $\bar{K}$ refer to the  center of a measured peak (with one detector).
Note that these quantities are determined from the kinematics of the neutron, in contrast
to $E_\textrm{int}$ which is a quantity of the scattering system.
Let us now try to  fulfill this equation with a pair $(E_\textrm{IA},K_\textrm{IA})$ being
determined by the impulse approximation for which holds  $E_\textrm{IA} = \hbar ^2 K_\textrm{IA}^2/2M$.
Then  Eq.~(\ref{E-conserv3}) yields
\begin{equation}
E_\textrm{IA} = \frac {\hbar^2  K_\textrm{IA}^2}{2 M_\textrm{eff}}  + E_\textrm{int} ,
\label{IA-assumption}
\end{equation}
where $ E_\textrm{int} >0 $  necessarily implies  $M_\textrm{eff} > M$.

From the considered increase of effective mass, \mbox{$M_\textrm{eff} >M$},
and Eq.~(\ref{IA-assumption}) one obtains the following experimentally testable predictions
of conventional theory:

$(a)$ Assuming  $K$-transfer to be fixed in a specifically designed experiment---a so-called
\emph{constant-}$K$ measurement---the recoil (i.e. kinetic) energy of an atom with mass
$M_\textrm{eff}$ will be \emph{smaller} than predicted by the impulse approximation (in which the
atom is free and has mass  $M$). An example is given in Fig.~2 of \cite{Watson}, which
demonstrates this effect with the aid of an exact calculation  of scattering from a harmonic
oscillator. The envelope of the exact transition lines is shifted to a \emph{lower} energy than
that of the calculated impulse approximation line; this shift corresponds to the recoil of a
fictitious free particle with a larger effective mass than  that of the oscillator.

$(b)$ Assuming $E$-transfer to be fixed in a specifically designed experiment---a so-called
\emph{constant-}$E$ measurement---the momentum transfer to a recoiling atom with mass
$M_\textrm{eff}$ will be \emph{larger} than predicted by  the impulse approximation (in which the
atom is free and has mass $M$).

It may be noted that various generalizations of the impulse approximation, including initial and final-state effects,
predict similar results. For example,~Stringari's well-known model \cite{Stringari} predicts a shift of
the measured recoil peak to lower $E$-transfer, in accordance with case $(a)$. Furthermore, this
shift was demonstrated with deep-inelastic neutron scattering data from liquid helium at $K = 10 \, \AA^{-1}$ \cite{Stringari}.

Summarizing,  conventional theory \emph{cannot} predict a
\emph{decrease} of the scatterer's effective mass---and/or
associated deviations of $E$- and $K$-transfers---opposite to
those of cases $(a)$ and $(b)$ .

\subsection{\label{sec:4.3}Neutron scattering -- Weak interaction}

As general (non-relativistic) scattering theory \cite{Taylor} shows,  scattering of an incoming
neutron wave packet
$\psi_n$ from a  particle
in the state  $\psi_A$ is described by
\begin{equation}
\psi_n \psi_A \rightarrow (\psi_n \psi_A)_\textrm{unscatt} + \varepsilon \Psi(n,A)_\textrm{scatt}
\label{scatt-theory}
\end{equation}
(up to normalization; $|\varepsilon| \ll 1$)
where the two subscripts denote the non-scattered and scattered components.
This process is represented  by an ordinary unitary transformation.
In the center-of-mass coordinate system, the scattered component is given by an outgoing
spherical wave (which is called $s$-wave scattering).  In particular, for neutron scattering it holds
\begin{equation}
\varepsilon \Psi(n,A)_\textrm{scatt} \propto  - \frac{b_M}{r} \exp(\imath k r) ,
\end{equation}
where $r$ denotes the relative spatial distance between the two particles,  ${b_M}$ is the scattering length of
the atomic nucleus with mass $M$,
and $k$ a proper wave
vector. For the (arbitrary) choice of the minus sign in  this expression, see \cite{Squires}.

The range of the nuclear forces is very short (of the order of $10^{-15}$ m = 1 fm), and the
associated scattering length $b_M$  in the Fermi
pseudo-potential, Eq.~(\ref{Fermi}), is of similar order, say 1--10 fm. Due to the much larger de Broglie
wavelength of the impinging  neutrons (being of the order of 1 \AA for thermal neurons), the
scattering  contains only $s$-wave components  \cite{Squires,Lovesey}.

Note that usually the struck nucleus (atom) is  not fully fixed and thus  exhibits
some recoil, which is particularly significant for a struck H-atom. Hence
neutron and scattering particle are \emph{entangled} in their  two-body final state.

Due to the smallness of scattering lengths of all nuclei \cite{Squires} and the
associated ultrashort range of the nuclear
forces, the neutron-nucleus scattering (commonly referred to as  neutron-atom scattering)
represents a weak interaction, and for this reason the theoretical framework of \emph{first-order}
perturbation theory is assumed to be fully sufficient.
In this context, it should be reminded that the general \emph{van Hove
formalism} of time-correlation functions
\cite{vanHove},
on which every conventional neutron
scattering theory is based (cf.~textbooks \cite{Squires,Lovesey}), contains  time-dependent
(Heisenberg) operators of the \emph{undisturbed} \mbox{$N$-body} scattering system only---i.e. the
disturbance caused by the neutron-system interaction is assumed  infinitesimally weak.

For further insight on how \emph{weak} the neutron-atom interaction is, the following fact
may be noted. Every existing neutron beam (e.g.~at the newest and most intense spallation source
SNS)    is \emph{weak} also in the sense that the probability for two neutrons of a
neutron pulse   to scatter off the same
atom  is practically zero. 

However, in the theoretical frame of  weak values and two-state-vector formalism, the weakness of
interactions   should be examined in some more detail. To yield a
numerical estimate, we take into account the discussion by Duck,
Stevenson and  Sudarshan    \cite{Duck1989}. These authors pointed
out that the weak value theoretical derivations of \cite{AAV1988}
require (among other conditions) that the state of the weak
measuring device must be representable by a wave function, rather
than by an impure density matrix. This can be achieved with the
probe beam being coherent across its width. 

In the neutron scattering experiments considered below (in Section~\ref{sec:6}),
the neutron's coherence width  is not specified in
the cited references. This is because this quantity plays no
role in conventional neutron scattering theory.
To determine this quantity, details of
the beam's monochromaticity and divergence would be  needed.  In
order to proceed, and as a rough estimate of the worst case, here
we may  assume this coherence width to be similar to, or larger
than, the neutron's wavelength $\lambda_n$. Thus the magnitude of
the total scattering amplitude of the neutron wave from a nucleus,
which should be proportional to $b_M$, can be roughly estimated by
the ratio
\begin{equation}
\varepsilon \lesssim \frac{b_M}{ \lambda_n} \sim \frac{10 \ \textrm{fm}}{ 1 \ \textrm{\AA}}
=   10^{-4} .
\label{smallness}
\end{equation}
Depending on  details of the instrumental setup, usually the
neutron-beam coherence width in real experiments should be larger than the neutron wavelength.
(It
may be noted that in various experimental setups of neutron \emph{interferometry}, this coherence
width can be of the order of 1 cm.) 

This crude estimate shows that the  non-relativistic scattering of  neutrons at issue can be
safely considered as a weak interaction process. Due to the smallness of the interaction range and
the ultrashort collisional time, the process is impulsive.

\newpage   
\section{\label{sec:5}Elementary scattering in the light of weak measurement and two-state-vector formalism}

\subsection{\label{sec:5.1}On shift operator and momentum transfer in impulsive two-body collisions
(conventional theory)}

In this section, the
position and momentum of the neutron (probe particle) are denoted as $(q,p) $. Similarly, the
position and  momentum of the scatterer (atom, nucleus) are denoted as $(Q,P)$. A symbol, say
$X$, with a hat, $\hat{X}$, represents the corresponding operator quantity. To illustrate the
action of shift operators, let us consider here a simple one-dimensional quantum model for
momentum exchange in a two-particle impulsive collision.

Let us furthermore assume that the two particles occupy states with approximately
well defined momenta (i.e. plane waves).
The initial state of the whole system is usually assumed to be uncorrelated:
\begin{equation}
\Psi_\textrm{initial}
=
\phi_{n}(p) \otimes  \Xi_{A}(P)
\label{initial state}
\end{equation}
(indices $n$ and $A$ refer to neutron and atom, respectively).

An impulsive scattering process  may be formally described by the (oversimplified) interaction
Hamiltonian
\begin{equation}
\hat{V}     =  F(t) \, ( \hat{q} - \hat{Q} ) ,
\end{equation}
where the function $ F(t) $ represents  a non-zero force  during a
short time interval $\tau$, i.e. the duration of the collision;
e.g.~we may assume  that $ F(t) $ is proportional to a delta
function.
Furthermore it is assumed that the integral  over $F(t)$
\begin{equation}
\int_0^\tau F(t) \, dt =  \hbar K
\end{equation}
gives the  momentum transfer $ \hbar K$ caused by the collision.

Neutron and atom observables commute, so
$ [ \hat{q} , \hat{Q} ] =0$, and the associated
unitary evolution operator is
\begin{equation}
\hat{U}(\tau) = e^{ - (\imath/\hbar) \int \hat{V} dt }
=
e^{ - (\imath/\hbar)\, \hbar K \,  (\hat{q} - \hat{Q}) }
\equiv
e^{- \imath \, K \, \hat{q}} e^{+ \imath \, K \, \hat{Q}} .
\end{equation}
The operator $ e^{ \imath \hbar K \,\hat{Q}/\hbar } $ shifts an atomic momentum eigenket as
$e^{\imath \hbar K \, \hat{Q} / \hbar} | P \rangle = | P + \hbar K \rangle $
while the operator
$e^{-\imath \hbar K \, \hat{q} / \hbar} $ shifts an impinging particle (neutron)
momentum eigenket
as $ e^{- \imath \hbar K \hat{q} / \hbar} | p \rangle = | p - \hbar K \rangle $.

Immediately after the momentum exchange, the state of the two-particle system in the momentum
representation is
\begin{eqnarray}
\Psi_\textrm{final} &=&
\hat{U}(\tau) \, \phi_{n}(p) \otimes \Xi_{A}(P)
\nonumber \\
&=& e^{-i \hbar K \, \hat{q} / \hbar}  \phi_{n}(p)
\otimes
e^{ i \hbar K \, \hat{Q} / \hbar}  \Xi_{A} (P) \nonumber \\
&=& \phi_{n} (p + \hbar K) \otimes   \Xi_{A}(P - \hbar K) .
\label{conventional}
\end{eqnarray}
This final state is not entangled, due to the trivial form of
$\hat{V}$.
These  short considerations  may motivate the search for a two-body
impulsive interaction Hamiltonian, which is presented in the
following section.

Concerning the basic importance of the shift operator in demonstrating
the dynamical non-locality of quantum mechanics, see the
discussion by Popescu \cite{Popescu-NatPhys}.

\subsection{\label{sec:5.2}Weak measurement, interaction Hamiltonian and momentum transfer}

As already mentioned in connection with Eq.~(\ref{Doppler}), for our purposes it is sufficient to
consider the atomic momentum component parallel to $\textbf{K}$, which from now on we shall denote
simply by $P$---and the associated operator  by $ \hat{P}$. In other words, in the following calculations
we shall consider the dynamics along an arbitrarily chosen direction
of momentum transfer, which is an one-dimensional problem.

\subsubsection{\label{sec:5.2.1}Von Neumann-type interaction Hamiltonian }

In this subsection, we provide a physically motivated heuristic
derivation of a von Neumann-type interaction
Hamiltonian for (a given amount of) momentum transfer in an impulsive collision.

Let us start again with the one-body model  Hamiltonian describing momentum transfer $-\hbar K\equiv +\hbar
K_n$   \emph{on the neutron} due to the collision with the atom:
\begin{equation}
\hat{V}_n(t)= \delta(t)\, \hbar K \, \hat{q} .
\end{equation}
As shown above, the associated evolution operator acting on the space of the neutron
\begin{equation}
\hat{U}_n(\tau) =
e^{ - \frac{\imath}{\hbar}\, \hbar K \,  \hat{q} }
\label{Un}
\end{equation}
shifts a   momentum eigenstate of the impinging particle (neutron) as $ e^{- \imath \hbar K
\hat{q} / \hbar} | p \rangle = | p - \hbar K \rangle $.

Assuming momentum conservation in the two-body collision, it holds
\begin{equation}
- \hbar  K_n = \hbar K_A \equiv  \hbar K ,
\label{K-definitions}
\end{equation}
where $\hbar K_A$ is the momentum transfer  \emph{on the atom} due to the collision.
(We choose $K_A$ with positive sign, following standard notation of conventional theory
\cite{Squires, Lovesey}.)

The scattering atom is assumed at rest in its initial state before collision, $\langle
\hat{P} \rangle_{i} = 0 $. After the collision, one conventionally expects that
\begin{equation}
\hbar K_A =+\hbar K = \langle \hat{P} \rangle_{f} =
\langle \hat{P}\rangle_{f} - \langle \hat{P}\rangle_{i}
\label{Pconventional}
\end{equation}
and correspondingly for the neutron momentum
\begin{equation}
\hbar K_n =-\hbar K =
\langle \hat{p}\rangle_{f} - \langle \hat{p}\rangle_{i} .
\label{Pconventional2}
\end{equation}
Hence, the aforementioned operator $\hat{U}_n(\tau)$ of the neutron, Eq.~(\ref{Un}),
may be written as
\begin{equation}
\hat{U}_n(\tau) =
e^{  - \frac{\imath}{\hbar}\, \langle \hat{P}\rangle_{f}
\,  \hat{q} } .
\label{U2}
\end{equation}

To apply the theory of weak measurement and two-state-vector formalism, a von Neumann two-body
interaction Hamiltonian is needed.
Thus one is intuitively guided to try a two-body generalization of the
one-body evolution operator $\hat{U}_n(\tau)$ of the form
\begin{equation}
\hat{U}(\tau) =
e^{ - \frac{\imath}{\hbar}\,  \hat{q} \, \hat{P}  } .
\label{U3}
\end{equation}
This heuristically obtained expression  still has  not  obvious context to the \emph{real}
experimental situations under consideration.
To  achieve this,  we now may proceed as follows.

Firstly, let us refer to the
aforementioned impulse approximation and Eq.~(\ref{E-conserv}) regarding energy conservation:
$$
E=  \frac {(\hbar K)^2}{2M} +\frac{\hbar{\mathbf K\cdot \mathbf P}}{M} .
$$
Looking at this equation, one sees that the \emph{larger} recoil term $\frac {(\hbar K)^2}{2M}$ may be
viewed to result from a \emph{strong} impulsive interaction (associated with momentum transfer $+\hbar
K$ on the atom). The theoretical treatment of this part of the interaction is \emph{not} within
the scope of the present paper. Since in the impulse approximation usually holds $|K| \gg |P|$,
the \emph{smaller} Doppler term $\frac{\hbar{\mathbf K\cdot \mathbf P}}{M}$ may correspond to a
\emph{weaker} interaction, in which the atomic momentum
$\hat{P}$
couples with  an appropriate dynamical variable of the neutron, say ${\cal{\hat{O}}}_n $. 
Looking  at the preceding formulas
(\ref{U2},\ref{U3}) for the model operator effectuating momentum transfer, it appears that this
dynamical variable should be  $\hat{q}$, that is ${\cal{\hat{O}}}_n  = \hat{q}$ .

Secondly, in view of the theory of weak values and two-sate-vector formalism, the  weak
interaction is expected to cause weak \emph{deviations} from  (or:  additional \emph{small}
contributions to) the conventionally expected \emph{large} momentum transfer $\hbar K$. This can be
introduced into the formalism by replacing $ \hat{P}$ with the \emph{small} momentum difference
$\hat{P} - \hbar K \hat{I}_A  $, and also including a positive \emph{smallness} factor
$$0<\lambda \ll 1$$
in the model interaction Hamiltonian and the associated evolution operator. In particular, let us
assume the model interaction Hamiltonian
\begin{equation}
\hat{H}_\textrm{int}(t) = + \lambda \,\delta (t) \,  \hat{q}
\otimes  (\hat{P} - \hbar K \,  \hat{I}_A ) .
\label{H-int}
\end{equation}

It may be noticed that this model Hamiltonian is not put forward entirely on kinematical grounds, like
Eq.~(\ref{K-definitions}), but also on quantum dynamical grounds, e.g. the choice \mbox{${\cal{\hat{O}}}_n  = \hat{q}$}
of the canonically conjugate operator $\hat{q}$ to neutron's momentum $\hat{p}$.

Moreover, it should be pointed out that the \emph{plus sign} in front of this expression is not arbitrary,
since it is consistent with the definitions (\ref{K-definitions}).
This point will be addressed explicitly in subsection \ref{sec:5.3.1}, since it plays a decisive role in the context of
the new quantum effect of momentum transfer deficit.

For further physical motivation of the two parts of the model Hamiltonian of Eq.~(\ref{H-int}), it
may be helpfully to compare the above reasoning with an example by Aharonov et
al.:
\begin{quote}
Consider, for example, an ensemble of electrons hitting a nucleus in a particle
collider. $\left[\ldots\right]$
The main interaction is purely electromagnetic, but there is also a relativistic and
spin-orbit correction in higher orders which can be manifested now in the form of a weak
interaction. \cite[p.~3]{AharonovEPJ2014}
\end{quote}

\subsubsection{\label{sec:5.2.2}Weak value of atomic momentum operator $\hat{P}$}

Let consider the atom as the system.
Since the weak value of the identity operator is
$ (\hat{I}_A )_w = 1$,
for the weak value of the  atomic coupling
operator  $\hat{P} - \hbar K\,\hat{I}_A $ in the above interaction Hamiltonian holds:
\begin{equation}
(\hat{P} - \hbar K \,  \hat{I}_A )_w = P_w - \hbar K .
\end{equation}
In the following, we first calculate the weak value $P_w$ of $\hat{P}$ for some characteristic (and
experimentally relevant)  final states in momentum space.
The results will reveal a striking
deviation---more precisely, a deficit---from the conventionally expected momentum transfer to the neutron; to the latter belongs
the pointer momentum variable $\hat{p}$ conjugated to $\hat{q}$ appearing in Eq.~(\ref{H-int}).

For the calculation of the weak value, it seems natural to use the momentum
space representation, as scattering experiments usually measure
momenta (rather than the positions of the scatterers in real space).

Let the atom initially be at rest and in a spatially confined state
(e.g. in a potential representing physisorption on a surface;
cf.~experiments  in Section~\ref{sec:6}). Then the initial atomic wave
function $\Xi(P)_i$ can often be approximated by  a Gaussian $G_A$
centered at zero momentum,
$$\Xi(P)_i \approx  G_A(P) .$$
At sufficiently deep temperature the atom will be in its ground state, and the width of $\Xi(P)_i$
is determined by the quantum uncertainty.

The struck atom moves in the direction of momentum transfer $\hbar K_A = \hbar K$; therefore, to
simplify notations, in the following calculations $P$ represents the  atomic momentum along the
momentum transfer direction.

It will be
instructive to  consider the following three cases for the atomic final state:
\begin{description}
\item{$(A)$ The final state is a plane wave (has vanishing width
in momentum space)---as assumed in general conventional theory and the impulse approximation.
Here, the result of conventional
theory is reproduced.}
\item{$(B)$ The final state  has a \emph{small} but non-vanishing width in momentum space.}
\item{$(C)$ Initial and final states have the same width in momentum space.}
\end{description}

$(A)$ \emph{Plane wave as final state}.
As in the case of the usual impulse approximation
\cite{Watson} of Compton scattering from a single scatterer, let the
final state be a plane wave; that is the momentum wave function is a
delta function $\delta_A$  centered at the assumed transferred
momentum $\hbar K_A$,
\begin{equation}
\Xi(P)_f =  \delta_A(P-\hbar K_A) .
\end{equation}
The weak value of $\hat{P}$ follows straightforward:
\begin{eqnarray}
P_w &=& \frac{\bra{\Xi_f} \hat{P} \ket{\Xi_i}}{\braket{\Xi_f}{\Xi_i}}
\nonumber  \\
&=& \frac{\int dP\, \delta_A(P-\hbar K_A) \, P \,
\Xi(P)_i}
{\int dP\,  \delta_A(P-\hbar K_A)\, \Xi(P)_i}
\nonumber  \\
&=&  \frac{\hbar K_A \, \Xi(\hbar K_A)_i }
{\Xi(\hbar K_A)_i}
\nonumber   \\
&=& + \hbar K_A \equiv + \hbar K .
\label{Pw-IA}
\end{eqnarray}
(Recall the notations of Eq.~(\ref{K-definitions}).) Hence, the weak value of the system coupling operator
$(\hat{P} - \hbar K \, \hat{I}_A )$ is just zero,
\begin{equation}
(\hat{P} - \hbar K \,  \hat{I}_A )_w = P_w - \hbar K =0 .
\label{noanomaly}
\end{equation}
(According to standard quantum scattering theory, the scattered wave
may acquire an additional phase factor, say $e^{\imath \chi}$, which does not affect the
preceding result because this factor cancels out in the fractions of Eqs.~(\ref{Pw-IA}).)

This physically means that, in this case, the new theory yields \emph{no correction} to the
conventionally expected value of momentum transfer (i.e.~$- \hbar K $, by assumption) shown by the
pointer of the measuring device, namely:
\begin{equation}
\langle \hat{p}\rangle_f  -\langle \hat{p}\rangle_i = -\lambda \,(\hat{P} - \hbar K \,  \hat{I}_A )_w
= 0 .
\label{no-deficit}
\end{equation}
Thus, the result of Eq.~(\ref{noanomaly}) is consistent with
conventional theory of the impulse approximation (or, more generally, of incoherent neutron
scattering);
cf.~also Eq.~(\ref{conventional}).\\  

$(B)$ \emph{Final state with non-vanishing width}.
Relaxing and generalizing  the strict impulse approximation
assumption of case \emph{(A)}, let us make here the more
realistic assumption that the atomic final state is represented by a
\emph{widened} delta-like symmetric function $\Delta_A$   with finite
(small) width, which is centered again at  the conventionally
expected momentum transfer $ \hbar K_A$. That is,
\begin{equation}
\Xi(P)_f =  \Delta_A(P-\hbar K_A)
\end{equation}
for which holds again $ \bra{\Xi_f} \hat{P} \ket{\Xi_f} =\hbar K_A$.
For example, a Gaussian with (much) smaller
width than the initial state fulfils these  conditions. The weak value of the atomic
momentum operator is then
\begin{eqnarray}
P_w &=& \frac{\bra{\Xi_f} \hat{P} \ket{\Xi_i}}{\braket{\Xi_f}{\Xi_i}}
\nonumber  \\
&=& \frac{\int dP\, \Delta_A(P-\hbar K_A) \, P \,
\Xi(P)_i}
{\int dP\,  \Delta_A(P-\hbar K_A)\, \Xi(P)_i}
\nonumber  \\
&=& + \hbar K_A - \pi(\hbar K_A) \nonumber \\
&\equiv& + \hbar K - \pi(\hbar K_A) ,
\label{deltaK-B}
\end{eqnarray}
where the small  momentum correction  $ \pi(\hbar K_A)$  fulfills
\begin{equation}
0 <   \pi(\hbar K_A) .
\end{equation}
This can be  easily seen  as follows. In the nominator
$$\int dP\, \Delta_A(P-\hbar K_A) \, P \, \Xi(P)_i
$$
the factor $\Xi(P)_i$ gives more weight to the \emph{left} side (i.e.
for $P<\hbar K_A$)  of the symmetric peak $\Delta_A(P-\hbar K_A)$
than to the \emph{right} side; this leads to an average of $P$ being
smaller than the  central position $\hbar K_A$ of the  final state
$\Delta_A$.

Obviously, the correction term $\pi(\hbar K_A)$ vanishes in the plane wave (and  also the impulse)
approximation limit, as already  considered in the preceding case $(A)$;  i.e.
$$
\Delta_A\rightarrow \delta_A \ \  \Rightarrow \ \
\pi(\hbar K_A) \rightarrow 0 .
$$

Furthermore, we see that the \emph{finite width of the final state} causes a    \emph{reduction} of the
conventionally expected  momentum transfer $\hbar K$; namely
\begin{equation}
(\hat{P} - \hbar K \,  \hat{I}_A )_w = P_w - \hbar K = - \pi(\hbar K_A)
\label{anomaly}
\end{equation}
and consequently for the \emph{additional} shift of the meter pointer variable holds
\begin{eqnarray}
\langle \hat{p}\rangle_f -  \langle \hat{p}\rangle_i
&=&  -\lambda\, (\hat{P} - \hbar K \,  \hat{I}_A )_w  \nonumber \\
&=&  + \lambda\, \pi(\hbar K_A) .
\end{eqnarray}
Here, we made use of the general result given by Eq.~(\ref{ReWV}); see also
Eq.~(\ref{changed-sign}).\\

$(C)$ \phantomsection\label{5.2.2C}\emph{Final state with unchanged width}.
The collisional process is here assumed to be \emph{soft} enough in order to leave  the shape of the
initial state unchanged. In other words,   let the final state have the same shape as the initial
state, but be centered at the transferred momentum; i.e.
\begin{equation}
\Xi(P)_f =  \Xi(P-\hbar K_A)_i .
\end{equation}
The weak value of the atomic momentum operator is now
as follows:
\begin{eqnarray}
P_w &=& \frac{\bra{\Xi_f} \hat{P} \ket{\Xi_i}}{\braket{\Xi_f}{\Xi_i}}
\nonumber  \\
&=& \frac{\int dP\, \Xi(P-\hbar K_A)_i \, P \,
\Xi(P)_i}
{\int dP\,  \Xi(P-\hbar K_A)_i\, \Xi(P)_i}
\nonumber  \\
&=& + \frac{\hbar K_A}{2}=+ \frac{\hbar K}{2} .
\label{deltaK-C}
\end{eqnarray}
The last equality  follows immediately from $(a)$ the two symmetrically distributed $\Xi$
functions around their central position
$\bar{P} =\hbar K_A/2$
and $(b)$ the linear term  $P$ in the integral of the
nominator. It should be noted that this result does not depend on the
width of $\Xi$, as long as the two $\Xi$ functions are not orthogonal to each other.

In other words, the correction term of the momentum transfer is here
$\pi(\hbar K_A) =  \frac{\hbar K_A}{2}$.

This is a quite interesting result because it represents a
momentum-transfer deficit of 50\%; i.e. the scattered neutron
measures a momentum kick being only half of the conventionally
expected value. In more detail:
$$(\hat{P} - \hbar K \,  \hat{I}_A )_w =  +\frac{\hbar K_A}{2} -\hbar K = -\frac{\hbar K}{2}
$$
and applying Eq.~(\ref{ReWV}), or Eq.~(\ref{changed-sign}), one obtains for the correction
to the  shift of the meter pointer variable:
\begin{equation}
\langle \hat{p}\rangle_f -  \langle \hat{p}\rangle_i = -\lambda\, (\hat{P} - \hbar K \,  \hat{I}_A )_w
= + \lambda\,\frac{\hbar K}{2} .
\end{equation}
In subsection~\ref{sec:6.2}, we will present and discuss a recent striking experimental finding being
qualitatively comparable with this numerical result.\\

The rather \emph{general} scheme of the above derivations provides evidence that the new
effect under consideration is not a specific feature of the simple time-of-flight experiment at
issue, but a general one of any field of scattering physics, e.g.~relativistic scattering in
high-energy physics. The immediate implication is that the general theory of weak measurement and
two-state-vector formalism may  be relevant for a very broad range of modern scattering
experiments.

\subsection{\label{sec:5.3}Quantum momentum-transfer deficit}

As the above calculations show, in all cases under consideration the weak value  $P_w$ is real.
Hence for the calculation of the expectation value of the meter output variable $\hat{p}$, which
is the neutron's momentum (conjugate to $\hat{q}$), the general result of Eq.~(\ref{ReWV})
applies. Here, it should be  noted the change of sign between the usually applied interaction
Hamiltonian of Eq.~(\ref{interaction}), as e.g.~treated in \cite{AAV1988,Duck1989}, and the
presently used specific form (\ref{H-int}). Therefore, there is a corresponding change of sign in
the
result of Eq.~(\ref{ReWV}), namely
\begin{equation}
\langle \hat{p} \rangle_f - \langle \hat{p} \rangle_i=  -  g\, \textrm{Re} [A_w] .
\label{changed-sign}
\end{equation}

The preceding results of Eqs.~(\ref{deltaK-B}) and (\ref{deltaK-C})
imply that the meter variable $\hat{p}$
will not take the conventionally expected value $-\hbar K = -\hbar K_A$, but an \emph{anomalous} value
due to the reduction term
$\lambda\,(\hat{P} - \hbar K \,  \hat{I}_A )_w = - \lambda\,\pi(\hbar K_A)$.
In particular, for the
\emph{total} momentum transfer shown by the pointer momentum we may write:
\begin{eqnarray}
\left[ \langle \hat{p}\rangle_f -  \langle \hat{p}\rangle_i \right]_\textrm{total} \ \
&=& \left[ \langle \hat{p}\rangle_f -  \langle \hat{p}\rangle_i \right]_\textrm{conventional} \nonumber \\
& &  \ \ \ \ \ \ \  + \left[ \langle \hat{p}\rangle_f -  \langle \hat{p}\rangle_i \right]_\textrm{correction} \nonumber \\
&=&  - \hbar K    + \lambda\,\pi(\hbar K_A). \ \
\label{WM-result}
\end{eqnarray}
This result represents the
new quantum effect of
\emph{quantum momentum-transfer deficit}:  the absolute value of  momentum transfer
on the neutron
predicted by the new theory is smaller
than that predicted by  conventional theory; in short,
$$
| - \hbar K  + \lambda\,\pi(\hbar K_A) | \leq | - \hbar K | .
$$

The weak measurement result (\ref{WM-result})  has some conceptual similarity with the result by
Aharonov et al.~discussed in Section~\ref{sec:3}, in which the interferometer mirror received momentum kicks
only from the insight, but the strange result was that these kicks (under the proper
post-selection) somehow succeed to \emph{pull it in} instead to push it out \cite{Aharonov2013}.

As mentioned above, the physical reason of this new effect is the specific quantum interference
being revealed by the theory at issue, which is associated with the post selection.
As far as we know, this effect has no explanation within the frame of conventional theory of
scattering. Moreover, it is in blatant contradiction with conventional expectations of neutron scattering theory
even qualitatively, as discussed in subsection~\ref{sec:4.2}.

\subsubsection{\label{sec:5.3.1}The positive sign of the model interaction Hamiltonian}

It was pointed out above that the \emph{plus sign} in front of model interaction Hamiltonian $\hat{H}_\textrm{int}$ 
in Eq.~(\ref{H-int})  is necessary and not due to an arbitrary choice; see subsection~\ref{sec:5.2.1}.
This positive sign is critical to the presented  conclusions.

However, one might object this point and claim the following:
If one replaces $+\lambda$ with a negative factor, say $-\mu$ (where $\mu>0$), in the Hamiltonian
(\ref{H-int}), one may carry through the above derivations to arrive at a \emph{positive} momentum
transfer correction without any violation of momentum conservation.

Here, we will show that this objection is incorrect.  
First, the calculations of
weak value
$(\hat{P} - \hbar K \,  \hat{I}_A )_w = -\pi(\hbar K_A)$
remain unaffected, since
the operator does not contain the factor $\lambda$. Second, the replacement of the positive factor
$+\lambda$  with the negative factor $-\mu$,   yields the new Hamiltonian
\begin{equation}
\hat{H}_\textrm{int,new}(t) = - \mu \,\delta (t) \,  \hat{q}
\otimes  (\hat{P} - \hbar K \,  \hat{I}_A )
\label{H-int-new}
\end{equation}
and, at the same time, causes  a \emph{change of the sign} of the general theoretical result (\ref{changed-sign}),
which for the specific correction to the pointer shift (\ref{changed-sign}) under consideration reads
\begin{eqnarray}
\langle \hat{p}\rangle_f -  \langle \hat{p}\rangle_i
&=&  +\mu\,(\hat{P} - \hbar K \,  \hat{I}_A )_w  \nonumber \\
&=&  - \mu\, \pi(\hbar K_A) .
\end{eqnarray}
In particular, the correction term to the pointer momentum shift is now negative. This
\emph{increases} the absolute value of the total shift:
$$
| - \hbar K  - \mu\,\pi(\hbar K_A) | \geq | - \hbar K |
$$
and, therefore, this increase is  qualitatively in contrast to the \emph{decreased} weak value of atomic
momentum
$$
(\hat{P})_w = + \hbar K_A  -\pi(\hbar K_A) .
$$
(Recall that $\hbar K_A = \hbar K$, by definition.)

In simpler terms, using the new interaction Hamiltonian (\ref{H-int-new})
we arrive at an unphysical result: The weak value of the atomic momentum transfer is still reduced
(with respect to conventional theory), but the measuring apparatus shows an increased momentum
transfer. Obviously, this unphysical paradox is an artifact of the wrongly chosen negative sign in
the model Hamiltonian Eq.~(\ref{H-int-new})---and \emph{not} a weakness of the general theory of
weak values and two-state-vector formalism.  Namely, this paradox vanishes if the correct (i.e.
physically appropriate) model Hamiltonian of Eq.~(\ref{H-int}) is applied.

     Contradicting  the above explanations, one might insist that the choice of the negative factor $-\mu$
in the model Hamiltonian (\ref{H-int-new}) should be still legitimate. However, as shown, Eq.~(\ref{H-int-new}) implies that the 
measuring system (i.e.~the scattered neutron) exhibits an \emph{increased} momentum transfer---which is tantamount to
the aforementioned (in Section~\ref{sec:4}) expectations of \emph{conventional} theory, in particular the conventional final-state effects. 
But the recent experimental results presented in Section~\ref{sec:6} 
are in blatant contrast to conventional theory.

The above considerations refer explicitly to both quantum systems (i.e. atom and neutron). This is because 
the model  interaction Hamiltonian $\hat{H}_\textrm{int}$ 
in Eq.~(\ref{H-int}) contains dynamical variables of both systems, and  therefore its physical significance cannot be 
discussed by considering the scattering atom (and the weak value of $\hat{P}$) only. 

\section{\label{sec:6}Experimental context and results}

In this section, the obtained weak measurement results are compared with real experiments.
The derivations of the preceding section should apply to both neutron scattering subfields of
interest---neutron Compton scattering and incoherent inelastic neuron scattering---as they
do not contain any specific assumption being valid in
one subfield only.
The presented experimental results may be considered as possible examples of the
theoretical analysis of Section~\ref{sec:5}; other explanations are presently unknown, but they cannot be excluded yet.

\subsection{\label{sec:6.1}Neutron Compton scattering from H atoms of a solid polymer}

\begin{figure}[t]
\begin{center}
\includegraphics[width=80 mm]{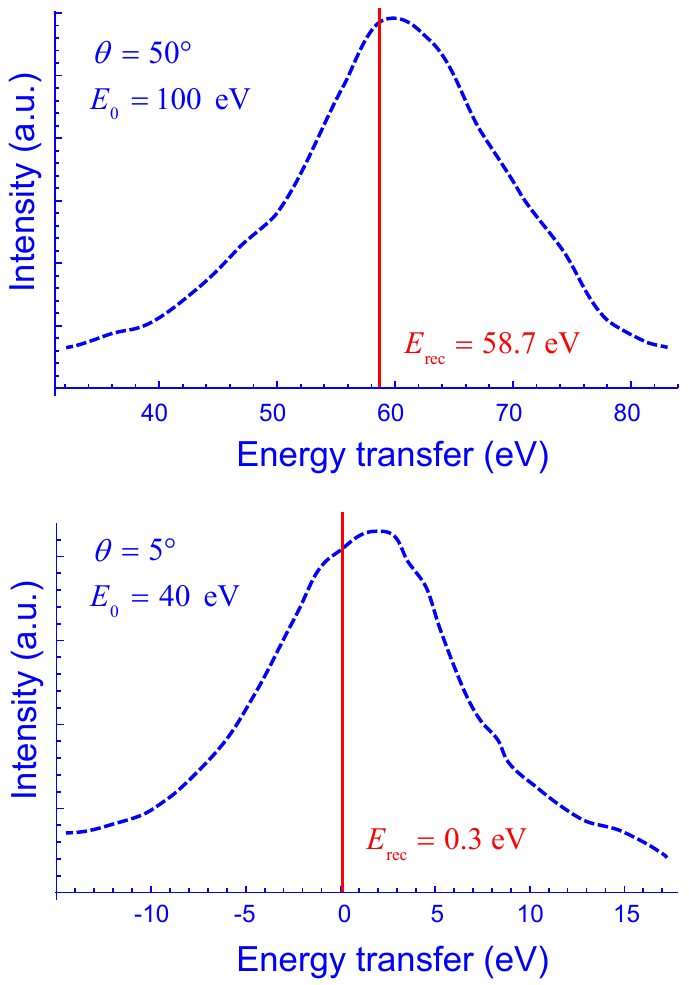}
\end{center}
\caption{\color[HTML]{0000FF}{\label{fig:3} Schematic representation of two examples of deep-inelastic neutron scattering spectra from a
solid polymer
(polyethylene, $[- \textrm{CH}_2 -]_n $) measured with the time-of-flight
spectro\-me\-ter MARI of ISIS \cite{Cowley2010}. Peak-shifts to \emph{higher} energy
transfers than the conventionally expected recoil energy $E_\textrm{rec}$ (vertical red lines)
are clearly visible. This corresponds to a \emph{lower} effective
mass of the recoiling H; see the text. For experimental details and more examples, see
\cite{Cowley2010}.}}
\end{figure}

An experimental demonstration of the new quantum effect under consideration can be found in
the data of \cite{Cowley2010}.  The deep-inelastic neutron scattering experiments were carried out by Cowley and
collaborators with the time-of-flight spectrometer  MARI \cite{MARI} of the neutron spallation source ISIS
(Rutherford Appleton Laboratory, UK).
The sample (a foil of a solid polymer) was at room temperature.
Fig.~\ref{fig:3} shows two examples of the extensive
measurements reported in \cite{Cowley2010}. The
depicted
recoil  peaks are mainly due to scattering from protons (H atoms), due to the high
scattering cross-section of H.
The vertical (red) lines show the $E$-transfer positions of the peaks
according to conventional theory,
Eq.~(\ref{recoil}).  The centroids  of the measured recoil peaks
are markedly shifted to
\emph{higher} energy transfer than conventionally expected.
As discussed in subsection~\ref{sec:4.2},
this shift is equivalent to a \emph{smaller}
effective mass of the recoiling H atom.

One might object that the shown data contain an additional small contribution from the C recoil.
However this is located at smaller $E$-transfers than that of H, due to their mass difference.
Therefore the above qualitative conclusion remains unaffected.

It may be noted that the shown neutron Compton scattering  peaks are very broad and asymmetric,
which is due to the (very)
low resolution of the employed modified setup  of MARI \cite{Cowley2010}. This makes a
quantitative analysis to determine the peak-position impossible. Nevertheless, visual inspection
of the data shows that the centroids of the peaks are \emph{shifted to higher energy} roughly by
5-10 \% of the recoil energy,
which equivalently  means that  the effective mass
$M_\textrm{eff}$ of the recoiling H atoms is \emph{smaller} than $M_\textrm{H}=1.0079$ a.m.u. of a
free H
by the same percentage
\begin{equation}
M_\textrm{eff}(\textrm{H}) \approx  0.91 - 0.96 \ \ \textrm{a.m.u.}
\end{equation}

It is also interesting to look at the additional (about 50) spectral data   reported
in \cite{Cowley2010}.  Remarkably, all those peaks appeared to show always  positive
$E$-transfers, i.e.~with no scatter to negative $E$-transfers;
see  \cite[Fig.~12]{Cowley2010}. This
underlines considerably the  reliability of the  $E$-shift  at issue.
However, since the main aim of that investigation
was  motivated differently (i.e. to measure the cross-section of H),
this striking experimental finding  remained fully
unnoticed in the discussions of \cite{Cowley2010}.
Noteworthy, throughout this section we presumed that the
\emph{calibration} of the instrument MARI employed in the experiment
\cite{Cowley2010} was correct, i.e.~that the applied time-of-flight method (see
subsection~\ref{sec:4.1}) used correct instrumental parameters.

\subsection{\label{sec:6.2}Incoherent inelastic scattering from single H$_2$ molecules in nanopores}

Another surprising result from incoherent inelastic neutron scattering was observed by Olsen et
al.~\cite{Olsen-H2} in the quantum excitation
spectrum of H$_2$  adsorbed in multi-walled nanoporous carbon (with
pore diameter about \mbox{8--20 \AA}).

The incoherent inelastic neutron scattering experiments were carried out at the new generation
time-of-flight spectrometer
of Spallation Neutron Source SNS (Oak Ridge Nat.~Lab., USA), called ARCS \cite{ARCS}. In this
experiment, the temperature was $T= 23$ K, and the incident neutron
energy $E_0$ was 90 meV.   The latter implies that the energy
transfer cannot excite molecular vibrations (or break the molecular
bond), but only excite rotation and translation (also called recoil)
of H$_2$ which interacts only weakly with the substrate:
\begin{equation}
E = E_\textrm{rot} + E_\textrm{trans} .
\label{rot-trans}
\end{equation}

\begin{figure}[t]
\begin{center}
\includegraphics[width=75 mm]{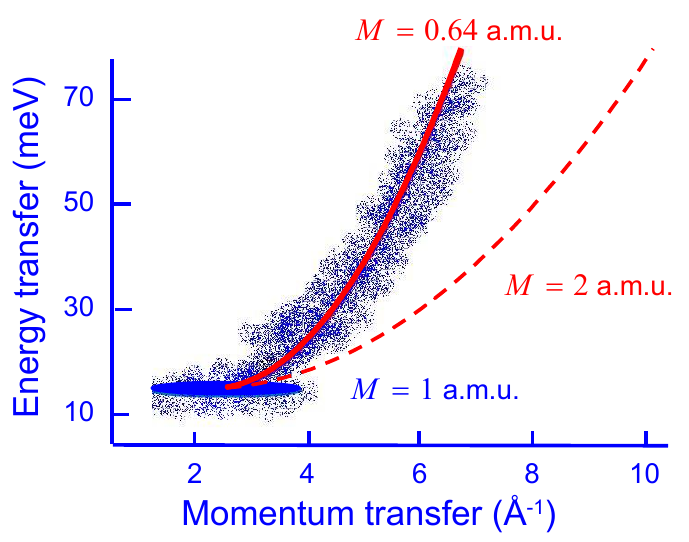}
\end{center}
\caption{\color[HTML]{0000FF}{\label{fig:4} Schematic representation of incoherent inelastic neutron scattering results
from H$_2$ in carbon nanotubes, with incident neutron energy $E_0 =90$ meV \cite[Fig.~1]{Olsen-H2}. 
The translation motion of the
recoiling H$_2$ molecules causes  the observed continuum of intensity, usually called \emph{roto-recoil} 
(white-blue ribbon),  starting at the well visible first rotational excitation
of H$_2$  being centered at $E \approx 14.7$ meV and $K \approx 2.7$ \AA$^{-1}$ (blue ellipsoid).
The $K-E$ position of the latter is in agreement with conventional theory.
In contrast, a detailed fit (red parabola; full line) to the roto-recoil data reveals a strong reduction of the
effective mass of recoiling H$_2$, which appears to be only 0.64 a.m.u. The red broken line (on the right) represents
the conventional-theoretical parabola with effective mass 2 a.m.u.
For details of data analysis see \cite{Olsen-H2}.}}
\end{figure}

The experimental two-dimensional incoherent inelastic neutron scattering intensity map $S(K,E)$ of H
(after background subtraction) is shown in Fig.~\ref{fig:4},
which is adapted  from the original paper \cite{Olsen-H2}. The
following features are clearly visible.
First, the  intensive peak  centered at
$E_\textrm{rot} \approx 14.7$ meV  corresponds to the well-known first rotational excitation
$J=0 \rightarrow 1$
of the H$_2$ molecule \cite{Mitchell-Buch}.
Furthermore, the wave vector transfer of this peak is  $K_\textrm{rot} \approx 2.7$ \AA$^{-1}$.
Thus the peak position in the $K$--$E$~plane shows  that the experimentally determined
mass of H   that  fulfills the relation $E_\textrm{rot} =(\hbar
K_\textrm{rot})^2/2M_\textrm{H} $ is (within experimental error) the mass
of the free H atom:
\begin{equation}
\textrm{rotation:}\ \ \ \ M_\textrm{H} = 1.0079\ \  \textrm{a.m.u.}
\label{M-rot}
\end{equation}
namely, ${M}_\textrm{eff}(\textrm{H}) = M_\textrm{H}$.
In other words, the location of this rotational excitation in the $K$--$E$~plane agrees with
conventional  theoretical expectations for incoherent inelastic neutron scattering,
according to which each  neutron scatters  from a single H \cite{Mitchell-Buch}.
Recall that an agreement with conventional theory was also observed  in the case of scattering
from $^4$He \cite{Glyde}; see Fig.~\ref{fig:2}.

Moreover,
the authors provide  a detailed  analysis of
the  roto-recoil data from incoherent inelastic neutron scattering, as shown in Fig.~\ref{fig:4}, and extracted a
strongly reduced   effective mass of the whole recoiling H$_2$ molecule (left parabola, red full line);
see Eq.~(\ref{recoil}):
\begin{equation}
\textrm{recoil:}\ \ \ \   {M}_\textrm{eff}(\textrm{H}_2) \approx 0.64  \pm 0.07 \ \ \textrm{a.m.u.}
\label{M-recoil}
\end{equation}
This is in blatant contrast to the conventionally expected value $M(\textrm{H}_2) =2.01$ a.m.u. for a
freely recoiling  H$_2$ molecule (right parabola, broken line). (Recall that the neutron-molecule
collision does not break the molecular H-H bond.)

An extensive numerical  analysis of the data is presented  in
\cite{Olsen-H2}, being based on
time-of-flight data analysis (cf.~Section~\ref{sec:4}) and the analysis of the
measured data within conventional theory
\cite{Squires,Mitchell-Buch}.

This strong reduction of effective mass, which is far beyond any
conceivable experimental error, corresponds to a strong reduction of
momentum transfer by the factor $~ 0.566$. Namely, the observed
momentum transfer deficit is about $ - 43\%$  of the conventionally
expected momentum transfer. This  provides first experimental
evidence of the new \emph{anomalous} effect of  momentum-transfer deficit
in an elementary neutron collision with a recoiling molecule.

Recall that, as explained above (see subsection~\ref{sec:4.2}), every H$_2$-substrate  binding must increase
the molecule's effective mass. Thus these findings from incoherent inelastic neutron scattering are in clear contrast to every conventional
(classical or quantum)  theoretical
expectation. However, they have a natural (albeit qualitative, at present) interpretation in the
frame of modern theory of weak values and two-state-vector formalism.

Incidentally, it may be noted that in principle the same \emph{calibration} of ARCS was used in
both experiments \cite{Glyde} and \cite{Olsen-H2}.

\subsubsection{\label{sec:6.2.1}Comparison with a related 1-dimensional experiment}

The  above experimental results  also show that the two-dimensional spectroscopic technique,
as offered by the  advanced time-of-flight spectrometer ARCS,  represents a
powerful method that  provides novel insights into quantum dynamics of molecules and condensed
matter. Clearly, this is due to the fact that  $K$ and $E$ transfers can be measured over a broad
region of the $K$--$E$~plane. This advantage makes these new instruments superior to the common
one-dimensional ones (like TOSCA at ISIS spallation source, UK), in which the detectors can only measure  along a
specific  trajectory in the $K$--$E$~plane. (TOSCA measures along two such trajectories
\cite{TOSCA}.)

As an example, consider the results of
\cite{Georgiev2005} from molecular H$_2$ adsorbed in single-wall carbon nanotubes   (which is
similar to the material of \cite{Olsen-H2}) at $T \approx 20$ K, investigated with TOSCA. Also this paper
reports the measurement of the roto-recoil spectrum, but as a function of $E$ only. Therefore the
strong anomalous effect  (\ref{M-recoil}) remained unnoticed, and for the theoretical
analysis of the data the mass of H$_2$ was
\emph{fixed}  to its conventionally expected value of 2 a.m.u.; see \cite[p.~903]{Georgiev2005}.

\subsubsection{\label{sec:6.2.2}Theoretical remarks}

The large value of observed
momentum  transfer deficit  in this experiment indicates that the
theoretical condition of \emph{weakness} is not fulfilled in this case.
Therefore, here we should mention the theoretical results by
Oreshkov and Brun  \cite{Oreshkov} which show that
weak measurements are universal, in the sense that
every generalized measurement can be decomposed into a \emph{sequence of weak measurements};
see remarks in subsection~\hyperref[7.2J]{7.2(J)}. This
implies that  the striking experimental result under consideration may still belong to the range
of applicability of the new theory of weak values and two-state-vector formalism.

In this context, a speculative semi-quantitative treatment of the effect's magnitude can be as
follows. A formal limit $\lambda \rightarrow 1$---which however is theoretically treated in
various works, e.g.~\cite{Qin2015}; cf.~subsection~\hyperref[7.2J]{7.2(J)}---taken in the above results shows that the special
case~\hyperref[5.2.2C]{$(C)$} predicts a $-50\%$ momentum transfer deficit (measured by the neutron), thus being
comparable with the experimental value of $-43\%$; see above. Additionally, the applied \emph{low}
excitation energy of 90~meV further supports the physical assumption that the excitation is
sufficiently \emph{soft} in order that   the envelope of the ground state
$\Xi(P)_i$  is  only slightly deformed, and thus the special case~\hyperref[5.2.2C]{$(C)$} applies.

It may be noted that this assumption about the final atomic state is very common in
the context of femtosecond dynamics in pump-probe optical
experiments on molecules:  A pump pulse lifts  the (Gaussian)
ground-state to an excited non-stationary state impulsively, keeping
its initial shape (or envelope) unchanged.

\subsection{\label{sec:6.3}Consequences for instrumental calibration}

The theoretical analysis and results presented above have  considerable implications for the
calibration of the associated time-of-flight spectrometers. In particular, in  neutron Compton scattering experiments it is a
common practice  to use the recoil peaks of certain light atoms (typically  He or H) to achieve a
\emph{refined} calibration
of the spectrometer. That is, the
measured $(K,E)$-positions of a peak---together with the standard free-atom recoil expression of
Eq.~(\ref{recoil}), sometimes also including conventional final-state effects---are used in order to
\emph{fine tune} the numerical values of (some of) the instrument parameters $( L_0,L_1^{\theta},t_0,\theta,
E_0 )$ determining the measured time-of-flight values, Eq.~(\ref{TOF}).
Obviously, such a calibration leads automatically
to an (artificial) agreement of the data with  conventional theory, thus being illegitimate for
testing the new predictions of weak measurement and two-state-vector formalism against those of conventional theory.

\section{\label{sec:7}Discussion}

\subsection{\label{sec:7.1}Reconsidering what is actually measured}

In view to the above  surprising experimental findings, it may be helpful to reconsider
them  with the aim to point out what is actually measured in the
experiment, and what is usually  assumed (implicitly or explicitly).

First of all, it should be reminded that all experimental results shown
in the figures of Section~\ref{sec:6} are in fact \emph{neutron} data, which are
\emph{converted} to H data with the aid of the conventionally assumed
theoretical relations
\begin{equation}
E_H=E_n \ \ \ \  \textrm{and} \ \ \ \ \hbar K_H = - \hbar K_n.
\label{conv-translation}
\end{equation}
These relations are
often considered as  \emph{trivial} or \emph{self-evident}, being (tacitly) based on the
assumption that the considered process is a two-body impulsive and incoherent (see Section~\ref{sec:4})
collision, with an additional assumption concerning the (effective) mass of the scattering
particle.

The \emph{experimental} data $(K_n,E_n)$ measured with the flight-of-time spectrometer
is to be compared with the differing (!) predictions of the
two alternative theories under consideration.

In the framework of the new theory (i.e. weak measurement and two-state-vector formalism), it is important
to realize that, in general, the impulsive collision creates an entangled \cite{Horodeckis}
(or a least discordant \cite{Modi-RMP2012}) neutron-H quantum state; see also
subsection~\hyperref[7.2F]{7.2(F)}.
Very shortly after the considered collision, the H-atom  becomes \emph{observed}, or \emph{measured}, by its close
environment (i.e. adjacent atoms), which of course changes the quantum-correlation content \cite{Modi-RMP2012}
of the neutron-H state. Later on, the (spatial position of the) scattered neutron becomes
\emph{strongly} measured by the neutron detector---which concludes the experiment.
In the context of the theoretical model of Section~\ref{sec:5}, taking into account the
the  environment of the
scattering H-atom,  we have:
\begin{equation}
E_\textrm{H+env}=E_n \ \ \ \  \textrm{and} \ \ \ \  \hbar K_\textrm{H+env} = - \hbar K_n
\label{H+env}
\end{equation}
assuming again energy and momentum conservation for the case that the
\emph{environment} of the scattering H is not neglected. Then,  $E_n=  E_\textrm{H+env}$
corresponds to
a momentum $\hbar K_A$ transferred on the scattering system ($A$ being here the system H+environment).
The theoretically derived momentum transfer deficit, when interpreted in terms of conventional theory,
is equivalent to a reduced effective mass of the whole H+environment system; i.e. the mass of the latter appears to be smaller than the mass of a free H-atom.

This weakness of conventional theory is further demonstrated by the results of the experiment by Olsen et al.~\cite{Olsen-H2},
as presented in subsection~\ref{sec:6.2}; see also subsection~\hyperref[7.2E]{7.2(E)}.
In contrast, the  contradictions
and/or  inconsistencies of conventional theory
just disappear in the framework of weak measurement  and two-state-vector formalism, and
at the same time, the experimental findings at issue are consistent
with the new theoretical  prediction of quantum momentum transfer deficit; see Section~\ref{sec:5}.
In particular, the experimental results appear to be due to a characteristic feature of weak measurement and two-state-vector
formalism  being unknown
in conventional theory, i.e. the specific quantum interference associated with
the \emph{finite width} of the  \emph{final} (i.e.~post-selected)
atomic state in momentum space and the basic formula of the \emph{weak value} of atomic momentum.

\subsection{\label{sec:7.2}Further remarks}

(A) Until now, weak values are measured with a different degree of
freedom of the \emph{same} quantum system.
Remarkably, in our approach, the two operators $\hat{q}$ and $\hat{P}$ occurring in the von
Neumann interaction Hamiltonian of Eq.~(\ref{H-int}) refer to two \emph{different} quantum
systems. Thus, according to Vaidman \cite{VaidmanComment2014}, the concept of weak value arises
here  due
to the
interference of a quantum \emph{entangled} wave and  therefore  it has no analog in classical wave
interference. This further supports
the conclusion   that weak value  is a genuinely quantum concept.\\

(B) To some people, \emph{post-selection} roughly means
``throwing some data out". In  the experimental context at issue,
however,
it rather means ``performing a concrete measurement on the system
at all, and analyzing the  measured data only".

Since the concepts of post-selection and associated quantum
interference play a central role in this study, let us stress the
following point. Strict energy conservation in the neutron-atom collision
(see Section~\ref{sec:4}, Eq.~(\ref{E-conserv})) reads
\begin{equation}
E \equiv    E_0-E_1  =
\frac {(\hbar \mathbf K + \mathbf P )^2}{2M} -\frac{P^2}{2M} .
\label{E-measurement}
\end{equation}
Let us consider the idealized  case in which the numerical values of  both neutron energies $E_i$
($i=0,1$) are sharply known---as conventional theory does by assuming \emph{plane waves} for the
neutron's initial and final states.
(As explained in Section~\ref{sec:4}, the detector
measures a time-of-flight value, from which
$E_1$ is derived if $E_0$ is pre-selected.)
Then, as  the initial atomic momentum ${\mathbf P}$ can
take a range of values around  ${\mathbf P} =0$ (due to the width of the atomic initial state
$\Xi({\mathbf P})_i$), the momentum transfer satisfying this equation can take various values too. In
other words, there are many   Feynman paths starting at $|E_0\rangle$  and ending at the
$|E_1 \rangle$, the latter being \emph{located} at the neutron detector. (These paths are interconnected with
associated atomic paths.) Thus these paths must be
\emph{coherently} superposed before the scattering probability of the neutron is calculated.
But as pointed out  above (in subsection~\ref{sec:4.1}), the formalism of conventional neutron
scattering theory fails to achieve this; e.g.~its basic result (\ref{S-IA}) for the impulse approximation contains
only the \emph{diagonal} part, Eq.~(\ref{n-IA}), of the initial-state atomic density operator
$(| \Xi \rangle\langle \Xi |)_i$, and no explicit  detail of the atomic final state.
In contrast, the theoretical approach of weak values and two-state-vector
formalism (Section~\ref{sec:5}) succeeds to achieve this goal.

In the frame of  two-state-vector formalism, however, and also making contact with Danan's et al.~paper
\cite{Vaidman2013}, the physical picture is substantially different from the conventional one. The
formulas of two-state-vector formalism imply that the scattering  under consideration is
determined by the overlapping  region---of momentum space, in our case---of both the forward-
and backward-evolving quantum waves of the whole neutron+atom system.
Concerning the atomic states, this overlap is obvious in the expressions of $P_w$.
Additionally, looking at
Eq.~(\ref{E-measurement}) we see that  also  the backward-evolving
neutron state $\langle E_1 |$ is consistent with a variety of atomic
momenta ${\mathbf P}$ and accompanying momentum transfers
$\hbar {\mathbf K}$.
Clearly, all associated Feynman paths connecting  $|E_0 \rangle$ with $|E_1 \rangle$
contribute to the considered neutron-atom collision.\\
      
(C) The derived quantum deficit of momentum transfer,  $- \pi(\hbar K_A)$,  bears some resemblance
to the theoretical result by Aharonov et al.~discussed in Section~\ref{sec:3},
in which an interferometer mirror was pulled inwards due to proper post-selection, despite the
photon momentum pushing it in the opposite direction \cite{Aharonov2013}.
	The innovation of the derived result in Eq.~(\ref{WM-result}) in comparison to the 
	theoretical results of 
Ref.~\cite{Aharonov2013}  is  the direct applicability of (\ref{WM-result}) to a widely known  subfield of
experimental neutron physics (and possibly also to further fields of scattering physics), 
as shown in Section~\ref{sec:6}.\\

(D) In the  physical context of the incoherent neutron scattering experiment of subsection~\ref{sec:6.1}, 
the   following remarks cannot be overemphasized:
Conventionally expected deviations from the impulse approximation, Eq.~(\ref{recoil}), (widely
known as final-state effects \cite{Watson}) must give peak shifts to less than $\hbar
\omega_\textrm{rec}$, since they are always caused by the atom not being free,  owing to its interaction
with other atoms. Thus there is an additional resistance to   motion of the struck atom, which
effectively \emph{increases} its mass and thus causes a slightly lower energy transfer than the
conventionally expected $\hbar \omega_\textrm{rec}$.
Summarizing, a
peak-maximum shift to higher energies than $\hbar \omega_\textrm{rec}$ seems impossible within
conventional theory; see subsection~\ref{sec:4.2} and \cite{Watson,Sears,Stringari}.)
Hence the considered experimental results, and the additional ones reported
in \cite{Cowley2010} showing the same peak-shifts to higher energy transfers,
contradict conventional theoretical expectations even qualitatively.\\

(E) \phantomsection\label{7.2E}Moreover, the results of \cite{Olsen-H2} discussed in subsection~\ref{sec:6.2} 
appear contradictory---in the light of conventional theory---because of the following two main points:

$(1)$ The observed  $J=0 \rightarrow 1$  rotational excitation of the H$_2$ molecule by
incoherent scattering exhibits an effective mass of $\ \approx 1$ a.m.u., Eq.~(\ref{M-rot}), as
conventionally
expected, since the neutron may be expected to exchange energy and momentum with  a single H.

$(2)$ However, in the same experiment,  the effective mass of the observed roto-recoil response
of the whole H$_2$ molecule is not 2 a.m.u. (because the whole molecule undergoes a translational
motion), but only $\approx 0.64$ a.m.u., Eq.~(\ref{M-recoil})---which is clearly meaningless
in the frame of   conventional theory.

As explained above, in conventional theory, depending on the specific constrains of measurement, a
\emph{decrease} of effective mass
is equivalent to $(a)$
an \emph{anomalous} increased energy transfer, or equivalently $(b)$ a decreased momentum transfer. In
both its forms $(a)$ and $(b)$, the observed effect appears---if interpreted within the
framework of conventional scattering theory---to contradict the basic laws of energy and/or
impulse conservation. In contrast, in the theoretical frame of weak measurement, this
contradiction dissolves.

In this context, note also that
\emph{exotic mass and momentum} values of weak measurements have been already
mentioned in \cite[Section~G.3]{Aharonov2014}.    \\

(F) \phantomsection\label{7.2F}As emphasized above, neutron-atom (or neutron-nucleus)
\emph{entanglement} \cite{Horodeckis}
or \emph{quantum discord} \cite{Modi-RMP2012},
caused by their scattering interaction, play
absolutely no role in conventional theory of neutron scattering \cite{Squires,Lovesey,Watson}.

Moreover, in neutron Compton scattering a second approximation is introduced by hand: The complete $N$-body
Hamiltonian of the scattering system is replaced with an  effective Hamiltonian of one particle
being captured in some effective \emph{Born--Oppenheimer} potential \cite{Watson, Tietje2011}.
Obviously this assumption is
tantamount to considering all interparticle entanglement or quantum discord (and the associated
decoherence appearing within the collisional time-window \cite{Stig-Aris}) as being irrelevant.\\

(G) Considerable efforts have been made during the last 25 years to compare Born--Oppenheimer potentials
calculated with quantum-chemistry methods with associated quantities derived from
several neutron Compton scattering data sets;
cf.~\cite{Watson} and references therein. In the light of the present investigation, however,
these comparisons appear to have a rather approximative character, thus being less
quantitative than claimed in the related publications.

To be  more specific, let us mention the following details. The overwhelming part of neutron Compton scattering
(or deep inelastic neuron scattering) investigations are dealing with measurements of \emph{Compton profiles} (i.e.~the
distribution of initial atomic momenta along a given spatial direction), from which one may derive
(properties of) the associated single-particle Born--Oppenheimer potential. The preceding theoretical results,
however, strongly question  the validity of the standard method of data analysis for the following
reason.

As shown in Section~\ref{sec:5},
the weak measurement-theoretical deficit  $- \pi(\hbar K_A)$ should depend on the considered momentum transfer
$\hbar K_A$. This  should cause an \emph{anomalous} \emph{distortion} of the
conventionally expected Compton profile, which  conventional theory assumes to be an even function
for isotropic systems. Namely, each point of the lower-momentum Compton peak-wing should receive a
larger weak value-shift $\pi(\hbar K)$ than a corresponding point of the higher-momentum wing.
Consequently, the widely applied \emph{symmetrization} procedure  of the measured Compton profile  (to
remove partially deviations from the impulse approximation; see \cite{Watson})  should introduce  artificial features
to the  \emph{derived Compton profile} and \emph{derived Born--Oppenheimer potential}. Especially for scattering from H the
Compton peak-width is particularly large (with respect to the instrumental resolution) and thus
the considered distortion should be measurable.
(This remark does not apply to the data of Fig.~\ref{fig:3}; see subsection~\ref{sec:6.1}.)\\

(H) The  derivations of Section~\ref{sec:5} apply to an impulsive two-particle collision, e.g.~of a neutron
with   an initially localized atom at rest. Since the scattering at issue is \emph{incoherent}---i.e. in each collision a neutron exchanges energy and momentum
with one scattering particle only---they apply directly to condensed and/or interacting systems
(like those considered in Section~\ref{sec:6}), where the scattering atoms are parts of a
molecule, or are adsorbed on surfaces or  nanocavities, etc.
Thus a scattering atom interacts with adjacent particles
of its  \emph{environment}  which is tantamount to a spatial confinement.
The latter causes a finite width of the initial atomic state $\Xi_i$,
and to less extend of its \emph{final} state $\Xi_f$---both appearing
to play an  equally important role in the quantum interference
captured in the weak value-calculations of subsection~\ref{sec:5.2}.

Based on these physical considerations,  we may conclude that the dynamics of the scattering atom
should also depend on the dynamics of the environment (which in general is also a quantum system),
and that  the collisional process is therefore
not a  strict two-body dynamical problem,
as standardly assumed in the impulse approximation of conventional theory; see \cite{Watson,Tietje2011}.
Especially, it is satisfying that the theoretical frame of weak values and two-state-vector formalism takes
care of these finite
widths of the initial \emph{and} final atomic states (see subsection~\ref{sec:5.2.2}), and in particular
reveals the \emph{significance of the final-state width} which plays a minor
role in conventional  theory of impulse approximation \cite{Watson} and its  extensions.\\

(I) The  remarks of points (G) and (H) support the opinion that weak values are novel quantum
interference phenomena in which \emph{post-selection} plays a crucial role
\cite{VaidmanComment2014,Romito2015,ParksGray2012,Dressel2015,Qin2015}. This opinion gets
additional support by the \emph{quantum Cheshire cat} effect \cite{Cheshire2013,Cheshire2014},
which however has found a classical analogue in a recent  experiment with continuous light beams
\cite{Cheshire2015}. In this regard, see also the work by Di Lorenzo \cite{DiLorenzo2013} who
shows that the quantum Cheshire cat is a consequence of quantum interference, that it is present
also for intermediate-strength measurements, and that it is a rather common occurrence in
post-selected measurements.

Further support of the quantum  character of the
Cheshire cat effect is presented by the generalization  to the case
that the pre- and the post-selections are entangled with each other
\cite{Popescu-complete-cat}, in which case there is no classical
analogue to this new Cheshire cat scheme.\\

(J) \phantomsection\label{7.2J}Generalizing the weak measurement formalism,
Oreshkov and Brun have shown \cite{Oreshkov}
that weak measurements are universal, in the sense that
any generalized measurement can be decomposed into a sequence of weak measurements. This important
theoretical result is further supported by the work of Qin et al. \cite{{Qin2015}}, who
showed that the main weak measurement results can be extended to the realm of \emph{arbitrary}
measurement strength.

This finding may have important applications to further collisional experiments than  the incoherent  neutron
scattering  at issue.
For example, energy and momentum transfers of various particles are the primarily measured
quantities in relativistic (high-energy) collision experiments too. Due to their complexity, full
details of the measurement methods and detection systems are usually not described in the related
publications. However it seems natural to believe that the above theoretical investigations can be
extended to the \emph{relativistic} scattering regime. Thus one may wonder whether the \emph{anomalous}
decrease of effective mass derived above can affect related relativistic measurements.

For example, the mass of the celebrated \emph{Higgs boson} was measured in proton-proton collisions and recently
reported to be ca.~125 GeV \cite{Higgs}. At the same time,  various categories of data (corresponding to
various decay channels) can be post-selected in order to determine the Higgs boson mass, and thus it would
be interesting to examine whether there might exist  significant differences between the values of
the mass correspondingly
determined; cf.~Fig.~1 of \cite{Higgs} in which such a (small) difference is visible.\\

(K) As mentioned above, a discussion concerning \emph{interpretation} and/or \emph{physical meaning} of
the concepts of weak value and two-state-vector formalism is beyond the scope of this paper.
However, it may be helpful for non-specialized or skeptical readers to consider Mermin's related
descriptions of some examples and results of  weak values and two-state-vector formalism
\cite{Mermin2011} using a more common scientific language.\\

Concluding, we feel that the theoretical formalism of weak values, weak
measurement and two-state-vector formalism  (which is
really quantum) not only sheds new light on interpretational issues concerning fundamental quantum
theory but  it also offers a new guide for our intuition to predict,  plan, and also carry out new
experiments and reveal novel quantum effects.

\section*{Acknowledgements} I wish to thank the Editors for constructive critical remarks and elucidative comments.

\end{document}